\documentclass{article}\usepackage[]{graphicx}\usepackage[]{color}
\makeatletter
\def\maxwidth{ %
  \ifdim\Gin@nat@width>\linewidth
    \linewidth
  \else
    \Gin@nat@width
  \fi
}
\makeatother

\definecolor{fgcolor}{rgb}{0.345, 0.345, 0.345}

\usepackage{framed}
\makeatletter
 {\par\unskip\endMakeFramed%
 \at@end@of@kframe}
\makeatother

\definecolor{shadecolor}{rgb}{.97, .97, .97}
\definecolor{messagecolor}{rgb}{0, 0, 0}
\definecolor{warningcolor}{rgb}{1, 0, 1}
\definecolor{errorcolor}{rgb}{1, 0, 0}
\newenvironment{knitrout}{}{} 

\usepackage{alltt}

\usepackage{graphicx}
\usepackage{natbib}
\usepackage{hyperref}
\usepackage{amsmath, amsfonts}

\newtheorem{theorem}{Theorem}
\newtheorem{lemma}{Lemma}

\newcommand{\Expect}[1]{\mathbb{E}\left[ #1 \right]}
\newcommand{\Var}[1]{\mathbb{V}\left[ #1 \right]}
\newcommand{\ModelDist}[1]{P_{n, #1}}
\newcommand{\ModelDens}[1]{p_{n, #1}}
\newcommand{\ModelDim}{d}
\DeclareMathOperator*{\argmin}{argmin}
\IfFileExists{upquote.sty}{\usepackage{upquote}}{}
\begin{document}

\title{A Note on Simulation-Based Inference by Matching Random Features}
\author{Cosma Rohilla Shalizi\thanks{Departments of Statistics and of Machine Learning, Carnegie Mellon University, 5000 Forbes Avenue, Pittsburgh, PA 15213 {\em and} Santa Fe Institute, 1399 Hyde Park Road, Santa Fe, NM 87501}}
\date{16 November 2021\\ Last \LaTeX 'd \today}
\maketitle

\begin{abstract}
  We can, and should, do statistical inference on simulation models by
  adjusting the parameters in the simulation so that the values of {\em
    randomly chosen} functions of the simulation output match the values of
  those some functions calculated on the data.  Results from the ``state-space
  reconstruction'' or ``geometry from a time series'' literature in nonlinear
  dynamics indicate that just $2\ModelDim+1$ such functions will typically suffice to
  identify a model with a $\ModelDim$-dimensional parameter space.  Results from the
  ``random features'' literature in machine learning
  suggest that using random functions of the data can be an efficient
  replacement for using optimal functions.  In this preliminary,
  proof-of-concept note, I sketch some of the key results, and present
  numerical evidence about the new method's properties.  A separate,
  forthcoming manuscript will elaborate on theoretical and numerical
  details.
\end{abstract}

\clearpage

\tableofcontents

\clearpage

\section{Introduction}

For decades, scientists have increasingly expressed their ideas as generative,
simulate-able models of complex processes; these models aim to capture both
mechanisms at work in the world and also measurement processes.  This is good
science, but a statistical problem.

Even when the phenomenon being simulated is complex, and the model produces
high-dimensional output, the underlying parameter space of the model will often
have many fewer dimensions.  (It is arguably just this compression that makes
the models insightful.)  As one example among many, the intricate
spatial-network epidemic model of
\citet{Chang-et-al-mobility-networks-and-covid-19} has just three adjustable
parameters not fixed by direct measurement or background knowledge.  The data,
on the other hand, consists of daily time series over multiple cities and a
period of months; it's 630-dimensional.  Similar pairings of high-dimensional
data and complex but comparatively-few-parameter models are common in
astronomy, climatology, economics, evolutionary biology, and ecology, among
other fields.  All the distributions these models produce can really be fitted
into a low-dimensional space (three-dimensional, in the case of
\citet{Chang-et-al-mobility-networks-and-covid-19}).  Statistical
analysis for such models should thus focus on parametric inference.

This, however, is where the problems start.  It is very common for these models
to be easy to simulate but hard to calculate with.  They can be ``run forward''
to generate detailed simulated data sets at low computational cost, even for
big models with many latent details that do not show up in the final data.  By
changing the parameters of the generative model, we change the distribution of
simulated outcomes.  But the mapping from parameters to distributions is
complicated, and we can't usually calculate the implied distribution.  It is
not feasible to find the probability of a particular outcome as a function of
the parameters, i.e., it is not feasible to calculate the likelihood function.
This rules out using classical statistical techniques (Bayesian or frequentist)
which rely on the likelihood function.

When we try to connect our simulation models to our data, we want the
simulations to match the data {\em somehow}, but it's often unclear {\em what
  aspects} of the data the simulations should try to match, and what they
should ignore as noise.

These three aspects of combining sophisticated models and rich data have made
it clear we need methods for non-likelihood inference.  Up to now,
practitioners have pursued strategies where simulation models are tuned to
match summary statistics or features calculated from the data.  These summaries
have been carefully crafted to ensure that (i) they are easily calculated from
data; (ii) their expectation values change rapidly as the parameters of the
simulation are adjusted; and (iii) any given value of the summary statistics
implies a unique value of the parameters, and vice versa.  There is typically
only one summary statistic per parameter.  Checking whether summary statistics
have these three properties is usually a lot of work, and if one's first
attempt doesn't meet the criteria, one has to start over and try something
else.

This paper is about making simulation-based inference much simpler and more
automatic.  The goal is to replace the procedure of carefully selecting of a
very small number of summary statistics, instead using about twice as many {\em
  random} functions of the data.  Results in nonlinear dynamics say that
``typical'' smooth functions from a $\ModelDim$-dimensional set (e.g., a 3-dimensional
family of distributions) into a higher, $k$-dimensional space are smoothly
invertible once $k > 2\ModelDim$, so that the values of $2\ModelDim+1$ ``typical'' functions
should be enough to identify $\ModelDim$ parameters.  Results from machine learning
show that randomly selecting a small number of functions of a high-dimensional
space can convey almost as much information as optimal summary statistics.
Bringing these ideas together suggests we can estimate models with a
(comparatively) small number of parameters from high-dimensional data, by
matching a small number of random features of the data, {\em without} the step
of carefully crafting summaries.

\S \ref{sec:overview} gives a high-level overview of the idea of this paper, as
well as fixing the necessary notation, and giving forward references to
sections of the paper clarifying some aspects of the idea.  \S
\ref{sec:background} goes over the related work I'm drawing on, in
simulation-based inference (\S \ref{sec:simulation-based-inference}), in the
use of random features in machine learning (\S \ref{sec:random-features}), and
in embeddings in nonlinear dynamics and topology (\S \ref{sec:embedding}); \S
\ref{sec:usual-asymptotics} recalls some general results about extremum
estimators from asymptotic statistical theory for context.  \S \ref{sec:theory}
sketches the theory, while \S \ref{sec:proof-of-concept} gives proof-of-concept numerical experiments.

\section{Overview, Setting and Notation}
\label{sec:overview}

You, the scientist, want to study a process in the world, which has produced a
multi-dimensional data point $x_n$ in a sample space $\mathcal{X}_n$.  (Here
$n$ may be a sample size, duration of a time series, extent of a spatial field,
etc.)  You have a generative model of the process you think is at work, with
some unknown parameters, $\theta \in \Theta \subset \mathbb{R}^p$.  Each value
of $\theta$ leads to a distribution $\ModelDist{\theta}$ over $\mathcal{X}_n$,
with densities $\ModelDens{\theta}(x)$ (with respect to some convenient
reference measure).  You would like to estimate $\theta$, or test whether
$\theta=\theta_0$, or quantify the uncertainty in $\theta$. When you ask me for
help in doing this, my instinct as a statistician is to try to do all these
things using the log-likelihood function.  You then break the news that
calculating $\ModelDens{\theta}(x)$ is intractable.  But you can simulate the model
easily, at whatever value of $\theta$ I ask for, giving me, say,
$\tilde{X}(\theta)$.

The core idea of this paper is that I should now draw $k$ real-valued functions
$F_1, \ldots F_k$ at random from a distribution $D$ over a suitable class of functions
$\mathcal{F}$, picking the functions independently of each other and of the
data.  Applying these functions to the observed $x_n$ gives a vector $F(x_n)
\in \mathbb{R}^k$.  My {\bf random-feature estimate} of $\theta$ would then
be\footnote{Using the squared distance, rather than the distance, and the
  factor of $1/2$ don't change the estimate, but they simplify some later
  expressions.}
\begin{equation}
\hat{\theta} = \argmin_{\theta \in \mathbb{R}^k}{\frac{1}{2}\left\|F(x_n) - \ModelDist{\theta}F\right\|^2} \label{eqn:ideal-random-feature-estimator}
\end{equation}
Going forward, abbreviate
$\ModelDist{\theta} F$ by $\Phi(\theta)$.  Conditional on the choice of $F$,
this is just an extremum estimator, so the usual theory of such estimators
applies (see \S \ref{sec:usual-asymptotics}).  In particular, if (i) $\Phi(\theta): \mathbb{R}^d \mapsto
\mathbb{R}^k$ is smooth with a smooth inverse, and (ii) a concentration
property holds, so that $F(X_n) \rightarrow \Phi(\theta_0)$ as
$n\rightarrow\infty$, familiar arguments establish the consistency of
$\hat{\theta}$.  If the optimum
in Eq.\ \ref{eqn:ideal-random-feature-estimator} is well-behaved (an interior
minimum with non-singular Hessian, etc.), then conventional arguments establish
asymptotic standard errors.  If a central limit theorem holds for $F$, we get
asymptotic Gaussianity.

Since we generally can't calculate expectations exactly, $\Phi(\theta)$ may
seem as un-available as the likelihood.  But since you can simulate the model
easily, $\Phi(\theta)$ can be approximated by
$s^{-1}\sum_{r=1}^{s}{F(\tilde{X}^{(r)}(\theta))}$, where $r$ indexes
independent runs of the model.  An estimate based on such simulations will be a
{\bf simulation-based random-feature estimate}.

Readers familiar with the literature on simulation-based inference will have
recognized this as akin to simulated minimum-distance estimators such as the
``method of simulated generalized moments'' and ``indirect inference'', which
also involve matching summary statistics (see \S
\ref{sec:simulation-based-inference}).  Craft lore among practitioners going
back to the 1990s suggests that the summaries to be matched must be picked
carefully, as the wrong features will be uninformative about $\theta$.  If the
features are chosen well, though, lore suggests that $k=\ModelDim$ features are enough
to get consistency, asymptotic normality, etc.  The novelty in this paper lies
in using precisely $k=2\ModelDim+1$ features, drawn at random.

The idea that $2\ModelDim+1$ features should be enough comes out of nonlinear
dynamics, and ultimately topology.  Simplifying (see \S \ref{sec:embedding}),
``embedding'' a $\ModelDim$-dimensional manifold $\Theta$ into $\mathbb{R}^k$
means finding a function $f: \Theta \mapsto \mathbb{R}^k$ which is smooth,
invertible, and has a smooth inverse.  It turns out that as soon as $k \geq
2\ModelDim+1$, the ``typical'' $f$ is an embedding, so long as each coordinate
of $f$ is a $C^1$ function and $\Theta$ is compact.  Since $\Phi(\theta)$ is a
map from $\Theta$ to $\mathbb{R}^k$ which we would like to be an embedding,
$2\ModelDim+1$ features should ``typically'' suffice.  (\S \ref{sec:embedding}
will explain the meaning of ``typical'' here.)

The idea that {\em random} features should be almost as informative as
carefully-chosen features comes out of work on ``replacing optimization with
randomization'' in machine learning (\S \ref{sec:random-features}).  A
characteristic result in this literature, for instance, asserts that predictors
of the form $\sum_{i=1}^{k}{\alpha_i F_i(x)}$, for $F_i$ drawn iidly from a
fixed distribution over $\mathcal{F}$, will have nearly the same risk as
predictors of the form $\int{\alpha(\omega) F(x;\omega) d\omega}$.  Further,
such random functions can themselves serve as function bases, with powerful
approximation properties.  These results work especially when the inputs $x$
are themselves high dimensional.

Bringing these results about random features together with the results on
embedding {\em suggests} that using as few as $2\ModelDim+1$ random features
should be almost as good for estimation as even the optimal selection of
features.

\section{Background and Related Work}
\label{sec:background}

I will only highlight work which immediately inspired these ideas, or forms an
obvious alternative approach.

\subsection{The Usual Asymptotics}
\label{sec:usual-asymptotics}

It will help to recall some well-established results about extremum estimators
(as found in, e.g., \citet[vol.\ I]{Gourieroux-Monfort-stats} or \citet[\S\S 5.2 5.3, 5.6]{van-der-Vaart-asymptotic-stats}).  We have a
sequence of random loss functions $M_n(\theta)$ (which are $X_n$-measurable).
The estimator is $\hat{\theta} \equiv \argmin_{\theta \in
  \Theta}{M_n(\theta)}$.  Assume that as $n\rightarrow\infty$, $M_n(\theta)
\rightarrow m(\theta)$, where the non-random limiting function $m$ has a unique
minimum at the $\theta_0$ which generated the data.  Further assume that
$\theta_0$ is in the interior of $\Theta$, and that $M_n$ and $m$ are regular
enough to allow whatever operations we need.  Taylor-expanding the gradient
$\nabla M_n(\hat{\theta}) = 0$ around $\theta_0$ yields a sandwich covariance
matrix for $\hat{\theta}$:
\begin{eqnarray}
\Var{\hat{\theta}} & \approx & \left(\nabla\nabla m(\theta_0)\right)^{-1}\Var{ \nabla M_n(\theta_0)} \left(\nabla\nabla m(\theta_0)\right)^{-1}
\end{eqnarray}
Gaussian fluctuations of $M_n$ around $m$ (for large $n$) will usually
translate into Gaussian fluctuations of $\hat{\theta}$ around $\theta_0$.

Specialized to the situation of \S \ref{sec:overview} above,
\begin{eqnarray}
  M_n(\theta) & = & \frac{1}{2}\| \Phi(\theta) - F(X_n)\|^2\\
  m(\theta) & = & \frac{1}{2}\| \Phi(\theta) - \Phi(\theta_0)\|^2
\end{eqnarray}
and abbreviating
\begin{eqnarray}
  \mathbf{g} & \equiv & \nabla \Phi(\theta_0)\\
  v(n) & \equiv & \Var{F(X_n)}
\end{eqnarray}
we get
\begin{eqnarray}
\nabla M_n(\theta_0) & = & (\Phi(\theta_0) - F(X_n))^T \mathbf{g}\\
\Var{\nabla M_n(\theta_0)} & = & \mathbf{g}^T v(n) \mathbf{g}\\
\nabla \nabla m(\theta_0) & = & \mathbf{g}^T \mathbf{g}\\
\Var{\hat{\theta}} & \approx & \left(\mathbf{g}^T \mathbf{g}\right)^{-1} \mathbf{g}^T v(n) \mathbf{g} \left(\mathbf{g}^T \mathbf{g}\right)^{-1} \label{eqn:asymptotic-variance}
\end{eqnarray}
The last result is conditional on the choice of random features $F$.  All else
being equal, then, the bigger $\mathbf{g} = \nabla \Phi(\theta_0)$, the more
precise our estimates will be.  Of course, just multiplying the functions in $F$
by large constants won't help, because that would scale up both the derivatives
and the variances, canceling out exactly.

--- I have used the ordinary (squared) Euclidean norm $\|F(X_n) -
\Phi(\theta)\|^2 = (F(X_n) - \Phi(\theta))^T (F(X_n)-\Phi(\theta))$ for
algebraic simplicty.  We could instead use $(F(X_n) - \Phi(\theta))^T
\mathbf{w} (F(X_n) - \Phi(\theta))$ for any symmetric, positive-definite
$k\times k$ matrix $\mathbf{w}$, with the obvious change in asymptotic
variances.  Ideally, we'd use $\mathbf{w} = (\Var{F(X_n)})^{-1}$.  This
suggests a two-step estimation procedure, where we first use an unweighted norm
(or one with a crude set of weights) to get an initial estimate
$\tilde{\theta}$, then simulate from $\tilde{\theta}$ to approximate the
variance matrix of the features and minimize the weighted norm to get a more
precise estimate \citep{Gourieroux-Monfort-simulation}.

A useful variant on this idea, due (so far as I know) to
\citet{Wood-inference-for-nonlinear-dynamic}, is to act as though $F(X_n)$ had
a Gaussian distribution, with mean $\mu_n(\theta)=\Phi(\theta)$ and variance
matrix $\mathbf{\Sigma}_n(\theta)$, and maximize the resulting likelihood.
This may be approximately true if $F(X_n)$ obeys a central limit theorem, and,
even if not, serves to put more emphasis (as it were) on the dimensions of $F$
which {\em should} closely match $\Phi(\theta)$, and less emphasis on those
dimensions of $F$ which are intrinsically noisier.  I will refer to this idea
as the use of a {\bf Wood likelihood}, emphasizing that this is not the
likelihood of the data, but an approximate likelihood for the features $F$.

\subsection{Simulation-based inference}
\label{sec:simulation-based-inference}

Many simulation modelers still make purely qualitative comparisons of
simulation output to empirical data, essentially relying on a combination of
theory and prior knowledge to constrain the form and parameters of models, plus
the ability of experienced practitioners to (as it were) ``smell out''
mis-fits.  Such qualitative approaches are sometimes quite structured and
sophisticated \citep{Windrum-at-al-validation-of-agent-based,
  OSullivan-Perry-spatial-simulation}.  But these approaches are,
intrinsically, incapable of quantifying uncertainty.

Much of the impetus for frequentist simulation-based inference has come from
econometrics \citep{Gourieroux-Monfort-simulation} and quantitative biology.
Here the obstacles to using the likelihood arise partly from having many latent
variables which would need to be integrated over (as emphasized by
\citealt{Gourieroux-Monfort-simulation}), and, especially in biology, from the
sensitive dependence of the dynamics on initial conditions (emphasized by
\citealt{Wood-inference-for-nonlinear-dynamic}).  One response is to
approximate the likelihood by doing density estimates on simulations, a
strategy which is still being elaborated on \citep{Cranmer-et-al-frontier} but
faces basic curse-of-dimensionality issues.

My focus here is instead on likelihood-free strategies for simulation-based
estimation.  As mentioned above, the core idea shared by most such strategies
is to pick some summary statistics, and then tune the $\theta$ parameters until
summaries calculated on simulations match summaries calculated on the
data\footnote{A recent and intriguing exception is the ``approximation
  computation via odds ratio estimation'' of
  \citet{Dalmasso-Izbicki-Lee-ACORE}, which views the likelihood ratio test as
  a classification problem, and learns a good classifier from simulations.  The
  correspondence between testing and set estimation then gives confidence
  regions.  It would be very interesting to compare these confidence sets to
  those arising from quantifying the uncertainty around simulation-based point
  estimates.}.  Methods differ, largely, in the nature of the summary
statistics.

In symbols, the common idea begins with $k$ real-valued functions of the data,
$f_1, \ldots f_k$, collectively $f: \mathcal{X}_n \mapsto \mathbb{R}^k$.  This
induces a mapping $\phi: \Theta \mapsto \mathbb{R}^k$ by $\phi(\theta) =
\ModelDist{\theta} f$.  The ideal estimator would then be
\begin{equation}
\argmin_{\theta \in \Theta}{\frac{1}{2}\|f(X_n) - \phi(\theta)\|^2}
\end{equation}
perhaps replacing the squared Euclidean norm with a weighted version as
mentioned above.  Consistency requires $f(X_n) \rightarrow \phi(\theta_0)$ as
$n\rightarrow\infty$.  This would lead to the asymptotic variance given by Eq.\
\ref{eqn:asymptotic-variance} above.

The ``simulation-based'' part comes from replacing the unavailable
expectations with Monte Carlo approximations:
\begin{equation}
\hat{\theta} = \argmin_{\theta \in \Theta}{\frac{1}{2}\|f(X_n) - \overline{f}(\theta, n, s)\|^2}
\end{equation}
where
\begin{equation}
\overline{f}(\theta, n, s) = \frac{1}{s}\sum_{r=1}^{s}{f(\tilde{X}^{(r)}_n(\theta))} \label{eqn:simulated-summaries}
\end{equation}
i.e., an average of $f$ over $s$ independent simulations of the model with the
parameter set to $\theta$.  Consistency requires that if $X_n \sim
\ModelDist{\theta}$, then $f(X_n) \rightarrow \phi(\theta)$ as
$n\rightarrow\infty$.  Under such an assumption, using $\overline{f}(\theta, n,
s)$ instead of $\phi(\theta)$ inflates the variance of $\hat{\theta}$ by a
factor of $1+1/s$ compared to Eq. \ref{eqn:asymptotic-variance}.  Alternately,
just simulate from $\hat{\theta}$, repeat the estimation on the simulation
outputs, and take the variance of the re-estimates.

\subsubsection{The Method of Simulated (Generalized) Moments}
seems to have been the first instance of this general scheme
\citep{McFadden-method-of-simulated-moments, Lee-Ingram-simulation-estimation}.
The goal, unsurprisingly, was to approximate the celebrated ``generalized
method of moments'' of \citet{LPHansen-GMM}, where each $f_i$ function is
itself an average over suitable units (time points, spatial locations,
experimental subjects, etc.).  Making the $f_i$s be averages means that
laws of large numbers or ergodic theorems can be used to prove the convergence
$f(X_n) \rightarrow \phi(\theta_0)$.  Selecting the right generalized moments
usually involves either detailed inspection of the model, or treating each
coordinate of $X_n$ as a ``moment''.

\subsubsection{Indirect Inference (II)}

arose when \citet{Gourieroux-Monfort-Renault-indirect,
  Smith-first-paper-on-indirect-inference} moved away from relying on
\citeauthor{LPHansen-GMM}-style ``generalized moments''.  Rather, II introduces
an ``auxiliary'' model, parameterized by, say, $\beta$, which is itself
estimated by minimizing its own loss function.  It is this estimate of $\beta$
which plays the role of $F(X_n)$, and as usual $\theta$ is adjusted to minimize
the distance between $F(X_n)$ and $\overline{F}(\theta,n,s)$, the average
estimate of $\beta$ from $s$ runs of the simulation with parameter $\theta$.
(Using $s$ simulation per parameter value inflates the variance of the
indirect-inference estimator by a factor of $1+1/s$ as before.)  Consistency
requires that $F(X_n) \rightarrow b(\theta_0)$ and $\overline{F}(\theta,n,s)
\rightarrow b(\theta)$, for some non-random and invertible ``binding function''
$b(\theta)$, along with some minor regularity conditions\footnote{The weakest
  set of such regularity conditions known to me are those in
  \citep{Zhao-limit-order-book}.}  Beginning in econometrics where it is still
widely used \citep{Halbleib-et-al-on-indirect-inference}, indirect inference
has spread to ecology \citep{Wood-inference-for-nonlinear-dynamic,
  Kendall-et-al-dynamical-tests-of-mechanistic-hypotheses} and even sociology
\citep{Ciampaglia-ind-inf-for-social-simulation}.

The auxiliary models used in indirect inference need to be easily estimable,
and sensitive enough to the $\theta$ parameters that $b(\theta)$ is invertible;
ideally a smooth mapping with a smooth inverse.  For time series,
linear-Gaussian autoregressive or vector autoregressive models are often used
(e.g., \citealt{DeJong-Dave-structural-macro}); for spatial data,
linear-Gaussian conditional autoregressive models.  However, efficient
estimation often involves a lot of work to devise informative auxiliary models;
the closer the auxiliary model is to being well-specified, the more efficient
indirect inference estimates will be.  \label{ref:nickl}
\citet{Nickl-Potscher-indirect-inference} provided an elegant way around this,
by making the auxiliary model a nonparametric density estimate based on the
method of sieves.  Unfortunately, their results presume that the generative
model produces IID data, and it is far from clear how to generalize their
intricate construction to dependent data, which is where simulation-based
inference is most needed.
\citet{Carrella-et-al-indirect-inference-by-prediction} ingeniously suggested
selecting and weighting summary statistics (including auxiliary parameter
estimates) by simulating the model at a many random values of $\theta$,
calculating a large suite of candidate auxiliaries, and then doing a
regularized linear regression of $\theta$ on the candidate.  While this may be
faster than explicit optimization, it begs the question of where good summaries
come from in the first place.  It also (implicitly) uses a linear approximation
to the inverse binding function $b^{-1}(\beta)$, which will usually be quite
nonlinear.

\subsubsection{Approximate Bayesian Computation (ABC)}

is a related but distinct strategy for simulation-based inference, especially
widely used in evolutionary biology and ecology \citep{Beaumont-on-ABC} and
epidemiology \citep[\S 4.3]{epidemic-models-with-inference}.  ABC tries to
approximate the Bayesian posterior distributions $\pi(\theta|x_n)$.  The basic
form goes as follows.  Pick a (generally vector-valued) summary statistic $f$,
as before, and a ``tolerance'' $\delta$.  Find the empirical value of the
summary, $f(x_n)$.  Now draw a $\theta$ at random from the prior $\pi(\theta)$;
simulate $\tilde{X}(\theta)$ from the selected $\theta$; calculate
$f(\tilde{X}(\theta))$.  If $\|f(\tilde{X}(\theta)) - f(x_n)\| \leq \delta$,
accept the $\theta$ and add it to the posterior sample, otherwise, discard it;
repeat.  The collection of accepted samples then approximates the posterior
$\pi(\theta|x_n)$.  More precisely, it approximates $\pi(\theta|f(x_n))$, which
will induce some distortions unless $f$ is a sufficient statistic for $\theta$.
Further distortion is induced by the use of the tolerance $\delta$.  Popular in
practice, ABC's theoretical properties are an on-going object of study (e.g.,
\citealt{Frazier-et-al-asymptotic-properties-of-ABC}).

Again, the choice of summary statistics $f$ is usually seen as crucial.
Attempts at automated choice of summary statistics, such as
\citet{Barnes-et-al-considerate-ABC-statistic-selection}, have usually aimed to
pick the most-nearly-sufficient statistic (or combination of statistics) from a
pre-defined menu\footnote{E.g., in that paper, they greedily minimized the
  mutual information between $\theta$ and the full set of statistics
  conditional on the selected sub-set, using an information-theoretic
  characterization of sufficiency.}.
\citet{Fearnhead-Prangle-semi-automatic-ABC} showed how to get good summary
statistics from a knowledge of the posterior $\pi(\theta|x_n)$ --- which is
what ABC is supposed to find\footnote{They showed that the posterior
  expectation $\Expect{\theta|x} = \int{\theta \pi(d\theta|x)}$ would be a good
  summary vector.  To break the vicious circle, they proposed to approach this
  by first running ABC with arbitrary summaries to get a rough approximation
  $\tilde{\pi}(\theta|x)$ to $\pi(\theta|x)$, then drawing $\theta$s from
  $\tilde{\pi}(\theta|x)$ and simulating a $y$ from each $\theta$, and
  approximating $\Expect{\theta|X=y}$ by a linear regression of $\theta$ on a
  library of transformations of $y$ (cf.\
  \citealt{Carrella-et-al-indirect-inference-by-prediction}), and finally
  re-running ABC with the new summary features.}.
\citet{Vespe-weak-lensing,Vepse-thesis} gave a more nearly constructive
procedure by drawing values of $\theta$ from the prior, simulating $x$ from
each $\theta$, and then using diffusion maps
\citep{Coifman-Lafon-diffusion-maps} to embed the $(\theta, x)$ pairs in a
common space; the leading eigenfunctions of the diffusion operator provided the
summary features.  Because this is fundamentally a kernel-based approach, it
may be possible to connect it to the random features results described below
(\S \ref{sec:random-features}).


\subsection{Random Features in Machine Learning}
\label{sec:random-features}

Inspired largely by \citet{Rahimi-Recht-random-features}, researchers in
machine learning have explored the uses of randomly selected ``features'',
i.e., functions of the data, for prediction and other statistical tasks.
Originally this was seen as a way to approximate kernel-based predictors.  That
is, the goal was to approximate predictors of the form $\sum_{i=1}^{n}{\beta_i
  K(x, x_i)}$, with $K(\cdot, \cdot):
\mathcal{X}\times\mathcal{X}\mapsto\mathbb{R}$ being the kernel function, with
predictors of the form $\sum_{j=1}^{p}{\alpha_j f(x;\omega_j)}$, with the
$\omega_j$ being sampled from some suitable distribution over a function space
$\mathcal{F}$ related to the kernel.  In particular, the positive-definite
properties of kernels suggested using the (normalized) Fourier transform of $K$
as a distribution over trigonometric functions and sampling from it, leading to
random Fourier features.  Beyond making kernel methods more practical for large
computational problems, random Fourier features have been employed to measure
dependence between random variables
\citep{Lopez-Paz-et-al-randomized-dependence-coefficient}, test statistical
independence \citep{Zhang-et-al-large-scale-kernel-independence-testing} and
conditional independence \citep{Strobl-approximate-kernel-CI-tests}, do
two-sample testing \citep[\S
3.3]{Sutherland-Schneider-error-of-random-Fourier-features}, etc.  Experience,
and some theoretical results (e.g.,
\citealt{Honorio-Li-random-Fourier-features-are-dimension-independent}) suggest
that these methods can work especially well when the data space $\mathcal{X}$
is high dimensional.  Other sets of random features, not based on Fourier
transforms, have recently been advocated for tasks such as goodness-of-fit
testing and evaluating the quality of Monte Carlo output
\citep{Huggins-Mackey-random-feature-discrepancies}.  While I am don't know of
any results which directly use random features in this way for simulation-based
inference, \citet{Briol-et-al-MMD-for-generative-models} study simulation-based
inference by minimizing the maximum mean discrepancy, a kernel-based
discrepancy measure whose random-feature approximation is studied by
e.g. \citet{Sutherland-Schneider-error-of-random-Fourier-features}.

Going beyond the viewpoint of approximating kernels, however,
\citet{Rahimi-Recht-random-kitchen-sinks, Rahimi-Recht-uniform-approximation}
developed powerful results on the strength of random features as function bases
in their own right.  Omitting minor regularity condition,
\citet{Rahimi-Recht-random-kitchen-sinks} established that if we draw functions
$F_1, \ldots F_K$ iidly from a space $\mathcal{F}$ and fit a predictor of the
form $\sum_{i=1}^{k}{\alpha_i F_i(x)}$ to training data, its excess risk
compared to the optimal predictor of the form $\int_{\mathcal{F}}{\alpha(f)
  f(x) df}$ is $O(1/\sqrt{k})$ with high probability.  Or, again,
\citet{Rahimi-Recht-uniform-approximation} shows that an arbitrary integral
mixture over $\mathcal{F}$ can be uniformly approximated to $O(1/\sqrt{k})$
using $k$ random functions from $\mathcal{F}$; furthermore, such integral
mixtures are dense in a reproducing kernel Hilbert space.  Thus for instance
random Fourier features allow us to approximate every function in the RKHS
induced by the Gaussian kernel, an extremely rich space.

\label{ref:kulhavy} From a different direction, \citet[pp.\ 115--117, pp.\
123--125]{Kulhavy-recursive} outlined an approach to constructing sufficient
statistics for $\ModelDim$-parameter exponential families by applying
$\ModelDim$ different linear functionals to the log-likelihood function;
provided the functionals are (linearly) independent, the result is, in fact, a
sufficient statistic in 1-1 correspondence with the canonical sufficient
statistic.  For example, one can pick $\ModelDim+1$ points in the parameter space, say
$\theta_0, \ldots \theta_{\ModelDim}$, and take the $\ModelDim$ log density ratios
$\log{\ModelDens{\theta_i}(x)} - \log{\ModelDens{\theta_0}(x)}$ as the
sufficient statistics, provided these functions are linearly independent of
each other, which will generally be the case for random $\theta$s.  The result
needs the model to be an exponential family, or enveloped within an exponential
family, so it doesn't hold for most interesting simulation models, but it
illustrates how a small number of random features can be highly informative
about parametric families\footnote{\citealt{Kulhavy-recursive}'s text implies
  that his constructions extends an idea in
  \citet{Dynkin-necessary-and-sufficient}, but, not reading Russian and not
  finding a translation, I haven't checked just what is due to which author.
  --- \cite{Montanez-CRS-LICORS-cabinet} used this idea to
  cluster local predictive distributions in a non-parametric spatio-temporal
  forecasting problem, and found experimentally that $2\ModelDim+1$ density ratios
  worked best when there were $\ModelDim$ clusters.}.

\subsection{``Embedology''}
\label{sec:embedding}

Beginning with \citet{Geometry-from-a-time-series}, researchers in nonlinear
dynamics pursued a program of ``attractor reconstruction'' or ``geometry from a
time series'', as follows.  We observe a one-dimensional signal $y(t)$ which is
a function $r(s(t))$ of some higher-dimensional state $s(t)$.  We assume $s(t)$
evolves according to a smooth, deterministic dynamical system, so $s(t) =
\rho_t(s(0))$ for a suitable semi-group of smooth functions $\rho_t$ (i.e.,
$\rho_{t+h} = \rho_t \circ \rho_h = \rho_h \circ \rho_t$).  The attractor of
this dynamic is a set $\mathcal{A}$.  Since $s(t)$ is unobserved, we form the
``time-delay vector'' $u(t) = (y(t), y(t-\tau), \ldots y(t-(k-1)\tau)$ for some
choice of delay $\tau$ and number of lags $k$.  It is unsurprising that
$y(t+\tau)$ can be predicted {\em approximately} from $u(t)$, especially for
large $k$.  What was surprising was that, in the ``typical'' or ``generic''
situation, there is a {\em finite} $k$ above which $u(t+\tau) =
\psi_\tau(u(t))$ for a {\em deterministic} function $\psi_\tau$, which will
also fix $y(t+\tau)$, More exactly, not only is there a differentiable mapping
$\phi: \mathcal{A} \mapsto \mathbb{R}^k$ which takes $s(t)$ into $u(t)$, but
this mapping has a differentiable inverse, and $\psi_h = \phi \circ \rho_h
\circ \phi^{-1}$.  The time-delay-embedding space of $u(t)$s is thus
equivalent, ``up to a smooth change of coordinates'', to the underlying state
space of $s(t)$s --- typically.

The mathematical basis for this is a classic result in differential geometry,
the \citet{Whitney-embedding} embedding theorem .  This tells us that once $k
\geq 2\ModelDim+1$, the set of ``embeddings'' (differentiable, invertible maps
with differentiable inverses) forms an open, dense set in the set of $C^{1}$
functions from the $\ModelDim$-dimensional manifold $\mathcal{A}$ into
$\mathbb{R}^k$.  Embeddings are thus ``generic'' in the sense in which that
word is used in topology.  The extension of this result to dynamical systems,
the \citet{Takens-embedding} theorem, rested on showing this was still true
when the coordinate functions of the mapping took the form $r \circ
\rho_{\tau}$, for generic choices of $\rho$, $r$ and $\tau$.  This was the
foundation for a large body of applied work in nonlinear dynamics, as well as
attempts to refine the underlying embedding theorem.
\citet{Abarbanel-analysis, Kantz-Schreiber-2nd} review both theory and
applications.

For our purposes, the most useful paper from this field was
\citet{Sauer-et-al-embedology}.  This complemented the topological notion of
``typicality'' used by Whitney, Takens, etc., with a more probabilistic one.
What one wants to say is that ``almost every'' map is an embedding, but this
needs the intricate construction of a measure on an infinite-dimensional space.
Sauer et al. (Theorem 2.2) finessed this by say that maps are ``prevalent'':
given any smooth map $f: \mathcal{A} \mapsto \mathbb{R}^k$, $k \geq
2\ModelDim+1$, the perturbation $f+e$ is an embedding for Lebesgue-almost-all
maps $e$ in a finite-dimensional subspace $E$.  (If $\mathcal{A}$ is a subset
of $\mathbb{R}^l$, we can take $E$ to be the $l(2\ModelDim+1)$-dimensional
space of linear functions.)  This also implies that embeddings are dense among
the smooth mappings.

\subsection{Summary of Objectives}
\label{sec:agenda}

To sum up, the goal here is to establish that when we have a
$\ModelDim$-parameter generative model, we should be able to estimate the
$\ModelDim$ parameters by matching $k=O(\ModelDim)$ functions, chosen
independently from each other and from the data, drawn from a space of
functions which is both rich enough to include ``typical'' smooth functions,
and regular enough that sample values converge on expectations.  The hope is
that using randomly chosen functions will be (nearly) as efficient as using
optimally chosen functions, and a further hope is that we can get consistency,
and maybe even near efficiency, with just $k=2\ModelDim+1$.

\section{Theory}
\label{sec:theory}

The two main goals of this note are to introduce the idea of doing
simulation-based inference by matching random features, and to show that the
concept is feasible by means of numerical experiments.  A full exploration of
the theory is reserved for a separate manuscript in preparation.  Nonetheless,
it is worth making a few remarks.

Throughout, I will employ the following notation:
\begin{itemize}
  \item The parameters of the model live in $\Theta$.
  \item For each $\theta \in \Theta$ and $n \in \mathbb{N}$,
    $\ModelDist{\theta}$ is a probability measure on the sample space
    $\mathcal{X}_n$.  The family of probability measures so induced will be
    $\mathcal{P}_n$.
\end{itemize}
I will also make the following assumptions.
\begin{enumerate}
  \item $\Theta$ is a compact subset of $\mathbb{R}^\ModelDim$.
  \item The parameterization is non-redundant: if $\theta_1 \neq \theta_2$,
    then $\ModelDist{\theta_1} \neq \ModelDist{\theta_2}$.  (Actually this only
    needs to hold for all sufficiently large $n$.)
  \item For each $n$, the collection of probability measures
    $\mathcal{P}_n$ is a {\bf statistical manifold} in the sense of
    \citet{Diff-geo-in-stat-inf, Kass-Vos}.  In particular, if $\theta_m
    \rightarrow \theta$, then $\ModelDist{\theta_m} \rightarrow
    \ModelDist{\theta}$ in distribution, and vice versa.
\end{enumerate}

\subsection{Identifiability, Fisher-consistency and Consistency}

Let's say that a {\bf functional}, e.g., is a real-valued function of a
probability distribution, $\mathcal{P}_n \mapsto \mathbb{R}$.  Because, under
these assumptions, there is a smooth, 1-1 correspondence between the manifold
of probability measures and the parameter space, I will abuse notation a little
and also write $\Theta \mapsto \mathbb{R}$.  If we could work at this level,
that of functionals of probability measures / functions of the parameters,
there'd be little difficulty.  For $i \in 1:k$, say $\phi_i(\theta): \Theta
\mapsto \mathbb{R}$ are our $k$ distinct functionals, and $\phi(\theta): \Theta
\mapsto \mathbb{R}^k$ is the vector-valued functional we get by applying them
all at once (in parallel!).  Suppose that the individual $\phi_i$ are $C^1$ in
the parameters.  Then the embedding theorems (\S \ref{sec:embedding}) tell us
that, for large enough $k$, embeddings are an open, dense set in this space of
functions, and that Lebesgue-almost-all perturbations of non-embeddings are
embeddings.  If we just observed $\phi(\ModelDist(\theta_0))$, we could invert $\phi$ to recover $\theta_0$.  That is, Fisher-consistency would be typical,
once we used enough functionals which were $C^1$ in the parameters.

\begin{lemma}
  Assume all the things.  Let $\mathcal{M}$ be the collection of smooth,
  real-valued functionals on $\mathcal{P}_n$.  By composition, for any $m \in
  \mathcal{M}$, we may define a smooth function on $\Theta$.  Let $\mu$ be a
  collection of $2\ModelDim+1$ such functions, so $\mu: \Theta \mapsto
  \mathbb{R}^{2\ModelDim+1}$.  Then the set of $\mu$ which form embeddings of
  $\Theta$ in $\mathbb{R}^{2\ModelDim+1}$ is an open, dense set, and embeddings
  are prevalent.
\end{lemma}

\textsc{Proof}: Direct application of theorems 2.1 (genericity) and 2.2
(prevalence) in \citet{Sauer-et-al-embedology}. $\Box$

For better or for worse, we don't get to work with functions of the parameters.
We do not even really get to work with functionals of the probability measures,
since we only have {\em samples} from those measures.  The lemma is thus of
little direct use.  In order to make it useful, we need to relate it to
measurable functions of the data, i.e., to statistics.

Now, under the assumptions above, in particular under the statistical-manifold
assumption, as $\theta_m \rightarrow \theta$, then $\ModelDist{\theta_m} \rightarrow \ModelDist{\theta}$ weakly or in distribution.  This in turn means that for any bounded, continuous test function $f: \mathcal{X}_n \mapsto \mathbb{R}$, $\ModelDist{\theta_m} f \rightarrow \ModelDist{\theta} f$.  In fast, the topology of weak convergence is generated by neighborhoods of the form
\[
\left\{ P \in \mathcal{P}_n ~:~ |P f - r| \leq \delta \right\}
\]
varying $f$ over all bounded, continuous functions on $\mathcal{X}_n$, $r$ over
$\mathbb{R}$, and $\delta$ over $(0,\infty)$ \citep[p.\
260]{Dembo-Zeitouni-book}.  That is, the generating sets of the topology are
ones where the {\em expectation values} of bounded, continuous test functions
are close to target values.  It follows that smooth functionals on this
manifold are either the expectations of bounded, continuous test functions, or
are the limits of sequences of such expectations.  We thus have

\begin{lemma}
  Assume all the things.  Let $C_b$ be the collection of bounded,
  continuous functions on $\mathcal{X}_n$.  Let $f$ be a collection of $k$
  functions from $C_b$, and define $\phi_i: \Theta \mapsto \mathbb{R}$
  via $\phi_i(\theta) = \ModelDist{\theta} f_i$, and $\phi: \Theta \mapsto
  \mathbb{R}^k$ similarly.  Then the set of $\phi$ which are embeddings of
  $\Theta$ in $\mathbb{R}^{k}$ is an open, dense set, and embeddings are
  prevalent, once $k \geq 2\ModelDim+1$.
\end{lemma}

\textsc{Proof}: By the statistical-manifold assumption, each $\phi_i$ is a
$C^{1}$ function from $\Theta$ to $\mathbb{R}$.  Moreover, as discussed above,
the statistical manifold assumption also says that functions of this form are
dense in the space of $C^1$ functions of the parameters.  Now invoke the
previous lemma.  $\Box$.

Having whittled down from needing to get generic functionals of the
distribution to just generic expectations of bounded, continuous test
functions, it would be nice to say something about smaller classes
of test functions.

\begin{theorem}
  Assume all the things.  Let $\mathcal{F} \subset C_b$ be a class of bounded, continuous functions on $\mathcal{X}_n$, and suppose that linear combinations from $\mathcal{F}$ are dense in $C_b$.  Let $f$ be a collection of $k$ functions from $\mathcal{F}$, and define $\phi_i$ and $\phi$ as in the previous lemma.  Then the set of $\phi$ which are embeddings of $\Theta$ in $\mathbb{R}^k$ is an open, dense set, and embeddings are prevalent, once $k \geq 2\ModelDim+1$.
\end{theorem}

\textsc{Sketch}: This basically follows from the fact that the linear span of
$\mathcal{F}$ is, under these assumptions, weak-convergence-determining
\citep[p.\ 155]{Ethier-Kurtz-markov-processes}, and the previous lemma.

Notice that the random Fourier features have a very natural role here, as
convergence-determining class of bounded, continuous test functions.

The previous result is about what could be identified if we could work with
{\em expectation values} of generic continuous, bounded statistics.  Of course
we can't do that; we have only a single realization of the underlying process.
To make the transition from data to distributions, we need to ensure that
$f(X_{1:n}) \rightarrow \ModelDist{\theta_0} f$ as $n\rightarrow
\infty$, for all $f \in \mathcal{F}$.  The typical requirements for such a
concentration property are (i) some sort of Lipschitz condition, so that
$f(X_{1:n})$ doesn't depend too much on any one coordinate of $X_{1:n}$, and
(ii) some weak dependence conditions on the model
\citep{Kontorovich-Raginsky-concentration}.  This is basically all that's
needed; if the (idealized or infeasible) estimator $\argmin_{\theta}{\|f(X_n) -
  \ModelDist{\theta} f \|}$ would be consistent, then as a routine corollary
the simulation-based estimator $\argmin{\|f(X_n) -
  \overline{f}(\theta,n,s)]\|}$ will also be consistent, with the variance
increased by a factor of $1+1/s$ (p.\ \pageref{eqn:simulated-summaries}).

\subsection{Testing}

The whole previous development has focused on point estimation and its
uncertainty, but these ideas also apply to hypothesis testing and model
checking.  Testing the hypothesis that $\theta=\theta_0$ is as simple as seeing
whether $F(X_n)$ is acceptably close to $\Phi(\theta_0)$, with ``acceptably
close'' being determined by simulating from $\theta_0$.  Drawing additional
independent functions from $\mathcal{F}$, beyond those used to estimate
$\theta$, allows for goodness-of-fit.  (Cf.\
\citealt{Gelman-Shalizi-philosophy} on simulation-based model checking.)  If we
have a central limit theorem for the summary statistics, it would be possible
to supplement simulation-based tests with an asymptotic $\chi^2$ test.

\section{Numerical Experiments}
\label{sec:proof-of-concept}

The proof-of-concept is in the ability to actually run, so in this section, I
report some numerical experiments on random-feature-matching estimates for some
benchmark problems.

All numerical experiments were conducted using R, version 3.6.2 \citep{R}.  The
R file which generated all the figures and numerical results reported here is
available on request, though there are legal restraints on commercial use of
the code.

Throughout this section, I work with models which generate univariate sequences
or time series.  Experiments on higher-dimensional time series, and on spatial
and spatio-temporal processes, will be reported elsewhere.

I rely on random Fourier features, of the form
\begin{equation}
F_i(X_{1:n}) \equiv \frac{1}{n}\sum_{t=1}^{n}{\cos{(\Omega_i X_t + \alpha_i)}} \label{eqn:univariate-rffs}
\end{equation}
or
\begin{equation}
F_i(X_{1:n}) \equiv \frac{1}{n-1}\sum_{t=1}^{n-1}{\cos{(\Omega_i^{(1)} X_t + \Omega_i^{(2)} X_{t+1} + \alpha_i)}} \label{eqn:bivariate-rffs}
\end{equation}
Throughout, frequencies $\Omega \sim \mathcal{N}(0,1)$, and phases $\alpha$ were uniformly distributed on $(-\pi, \pi)$.
Random linear transformations were generated using the \texttt{expandFunctions}
package \citep{expandFunctions-package}.

The ``univariate'' random Fourier features of Eq.\ \ref{eqn:univariate-rffs} can be seen as empirical
counterparts to the characteristic function\footnote{Since ``characteristic
  function'' means different things in different fields: the characteristic
  function of a probability distribution $\mu$ over $\mathbb{R}^k$ is given by
  $\tilde{\mu}: \mathbb{R}^k \mapsto \mathbb{C}$ where $\tilde{\mu}(t) = \mu
  e^{i t\cdot X}$.  This is basically the Fourier transform, except for
  conventions about $i$ vs.\ $-i$.} for the marginal distribution of the $X_t$,
evaluated at the random frequencies $\Omega_1, \ldots \Omega_k$ (and with
random phase shifts).  The ``bivariate'' random Fourier features of Eq.\ \ref{eqn:bivariate-rffs} play the same role for the characteristic
function of the marginal distribution of the pairs $(X_t, X_{t+1})$.  Because
characteristic functions do, in fact, characterize probability distributions
\citep[Thm.\ 5.3, pp.\ 84--87]{Kallenberg-mod-prob}, the expectation values of
these features are good candidates for generic functionals.  Because these
features also time- or sample- averages, their sample values will converge on
expectations for a very broad range of data-generating processes.  Again,
further numerical experiments, using other random features, will be reported
elsewhere.

--- It would be natural to ask how random-feature estimates, in these examples,
compare to previous approaches to simulation-based inference.  Unfortunately,
this is not as straightforward as one would wish.  In the absence of truly
off-the-shelf implementations of those other techniques, I would have to craft
my own for each example.  But then there would always be the possibility that
(say) my implementation of indirect inference lost the race because I picked a
bad auxiliary model for some example.  The results of such comparisons will,
nonetheless, be reported elsewhere.

Because the objective function can be rather spiky (particularly for the
chaotic dynamical systems), they were optimized using generalized simulated
annealing, as implemented in the \texttt{GenSA} package \citep{GenSA-package}.
I limited each optimization to five seconds of computing time; some tinkering
(not included here) showed great improvements in going from $0.5$ to $2$
seconds, and again from $2$ seconds to $5$ seconds, but little improvement from
$5$ to $10$ seconds.  I have made no further effort to find the most
computationally-efficient way to minimize these objective functions.

\subsection{IID Process and Univariate Random Features}

\subsubsection{Gaussian Location Family}

\begin{figure}

\begin{knitrout}\small
\definecolor{shadecolor}{rgb}{1, 1, 1}\color{fgcolor}
\includegraphics[width=\maxwidth]{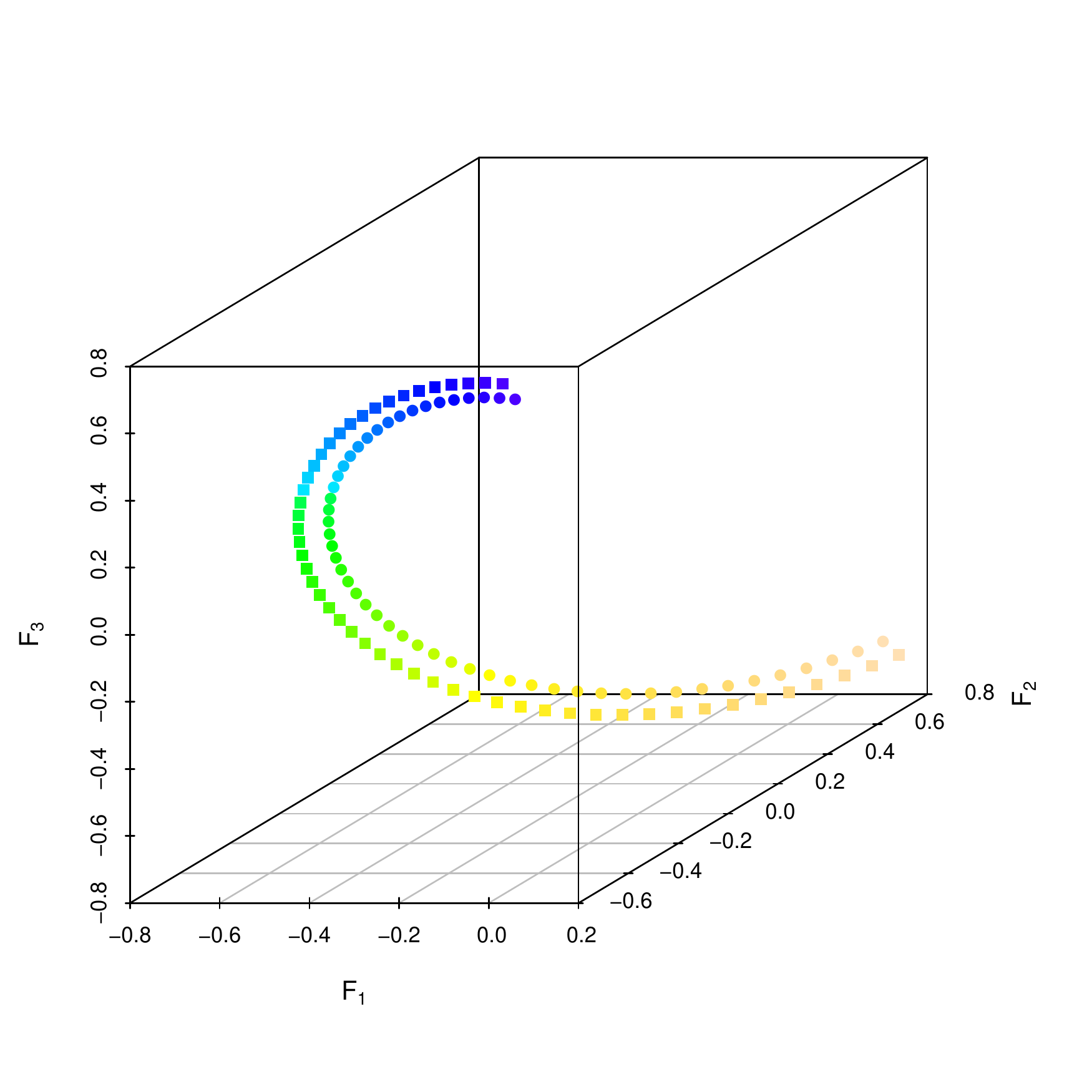} 

\end{knitrout}

\label{fig:univariate-random-Fourier-features-with-Gaussian-dist}
\caption{Using 3 random univariate Fourier features on the model $N(\theta,1)$,
  $n=100$, with feature values from 2 separate random realizations shown in
  two separate plotting symbols; color denotes $\theta$.  {\em Note:} This is
  using the ``French'' approach of using the parameters to transform a fix set
  of random draws, rather than re-randomizing at each parameter value.}
\end{figure}

I begin with the simplest possible baseline case, estimating the location $\mu
\in \mathbb{R}$ of an IID univariate Gaussian $\mathcal{N}(\mu, 1)$.  I begin
by checking that three ($=2\ModelDim+1$) random Fourier features {\em are}, in
fact, not only enough to uniquely identify $\mu$, but that the mapping is
smoothly invertible.  Figure
\ref{fig:univariate-random-Fourier-features-with-Gaussian-dist} shows that,
indeed, it is, and that sample-to-sample variation in the features
is already quite negligible at $n=100$.

\begin{figure}
\begin{knitrout}\small
\definecolor{shadecolor}{rgb}{1, 1, 1}\color{fgcolor}
\includegraphics[width=\maxwidth]{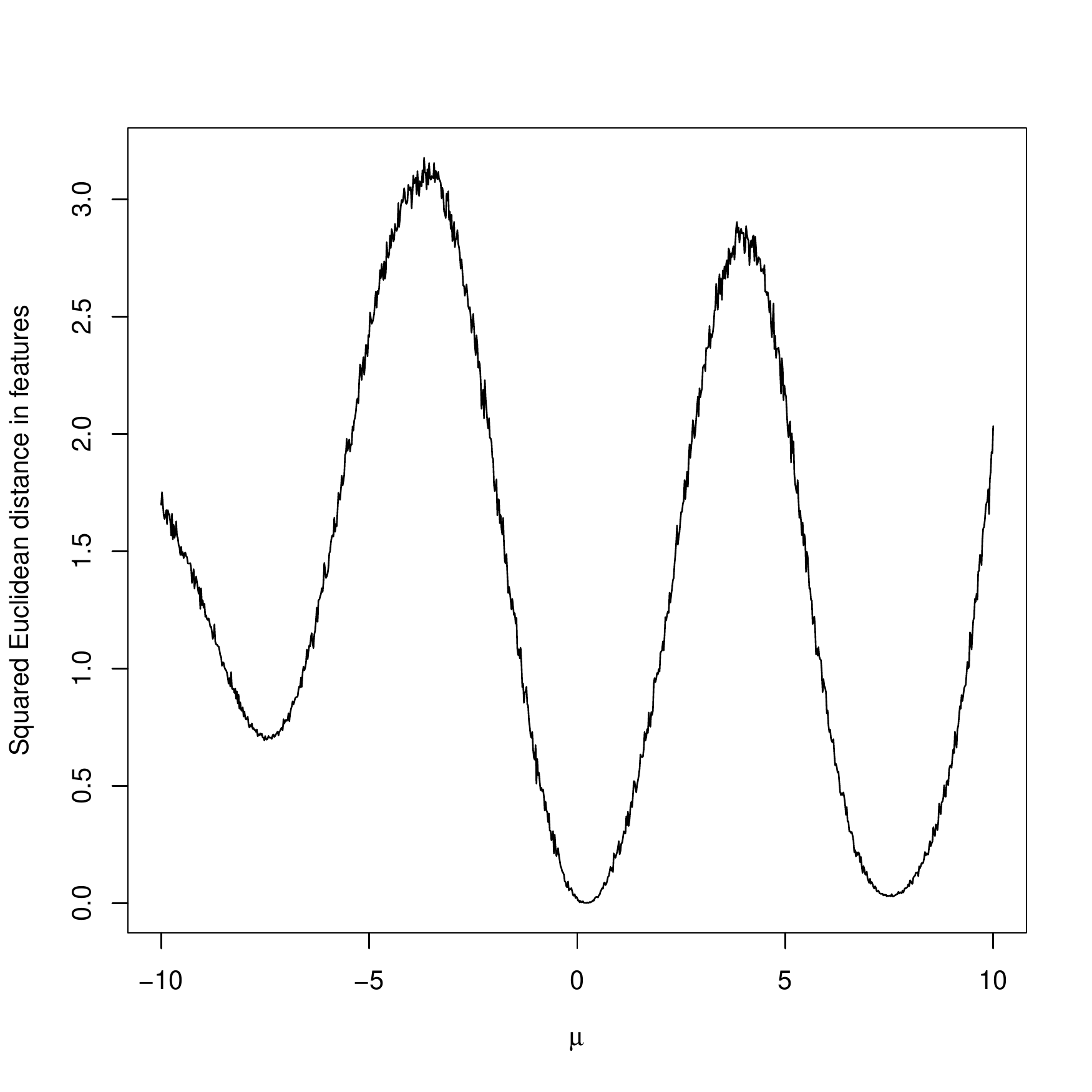} 

\end{knitrout}
\caption{Numerical evaluation of the distance between one sample of a standard Gaussian distribution ($\mathcal{N}(0,1)$), and unit-variance Gaussian distributions with varying means $\mu$.  The distance was evaluated using $10$ random draws per value of $\theta$, and using the same three random Fourier features used for the preceding plots.}
\end{figure}

\begin{figure}
\begin{knitrout}\small
\definecolor{shadecolor}{rgb}{1, 1, 1}\color{fgcolor}
\includegraphics[width=\maxwidth]{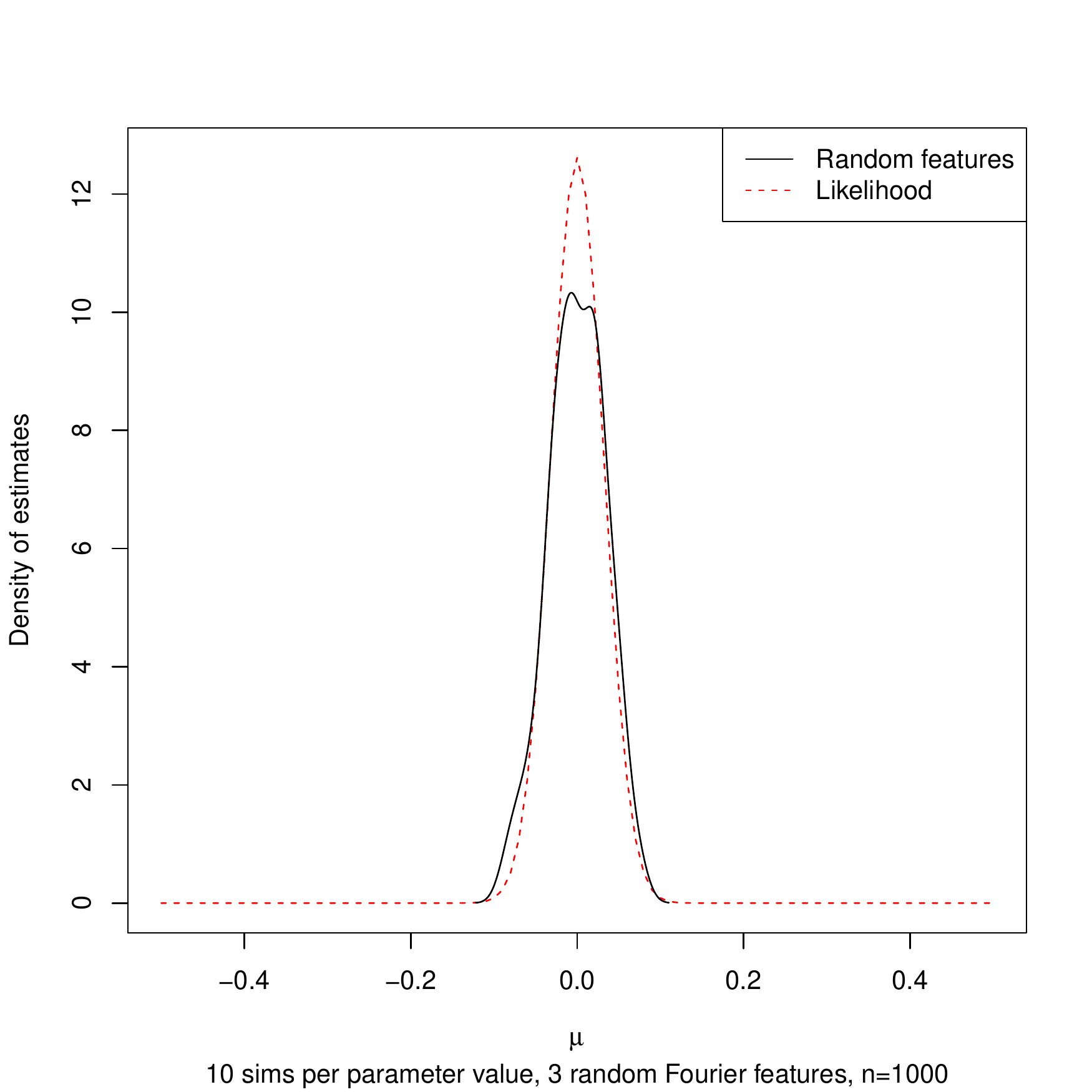} 

\end{knitrout}
\caption{Density of random-feature estimates (kernel-smoothed) versus the (theoretical) density of the MLE, when estimating the location of a univariate Gaussian by minimizing the distance with 3 random Fourier features.}
\label{fig:display.estimating.gaussia}
\end{figure}

The efficient way to estimate $\mu$ is, of course, just to use the sample mean,
which is the MLE.  The distribution of this estimator will be $\mathcal{N}(\mu,
1/n)$, and the MSE will be $1/n$.  Random feature matching will not do better
than this, but the question is how close it will come.  Figure
\ref{fig:display.estimating.gaussia} shows that we do, in fact, come rather
close, at least in this rather simple setting.

\clearpage

\clearpage



\clearpage

\begin{figure}
\begin{knitrout}\small
\definecolor{shadecolor}{rgb}{1, 1, 1}\color{fgcolor}
\includegraphics[width=\maxwidth]{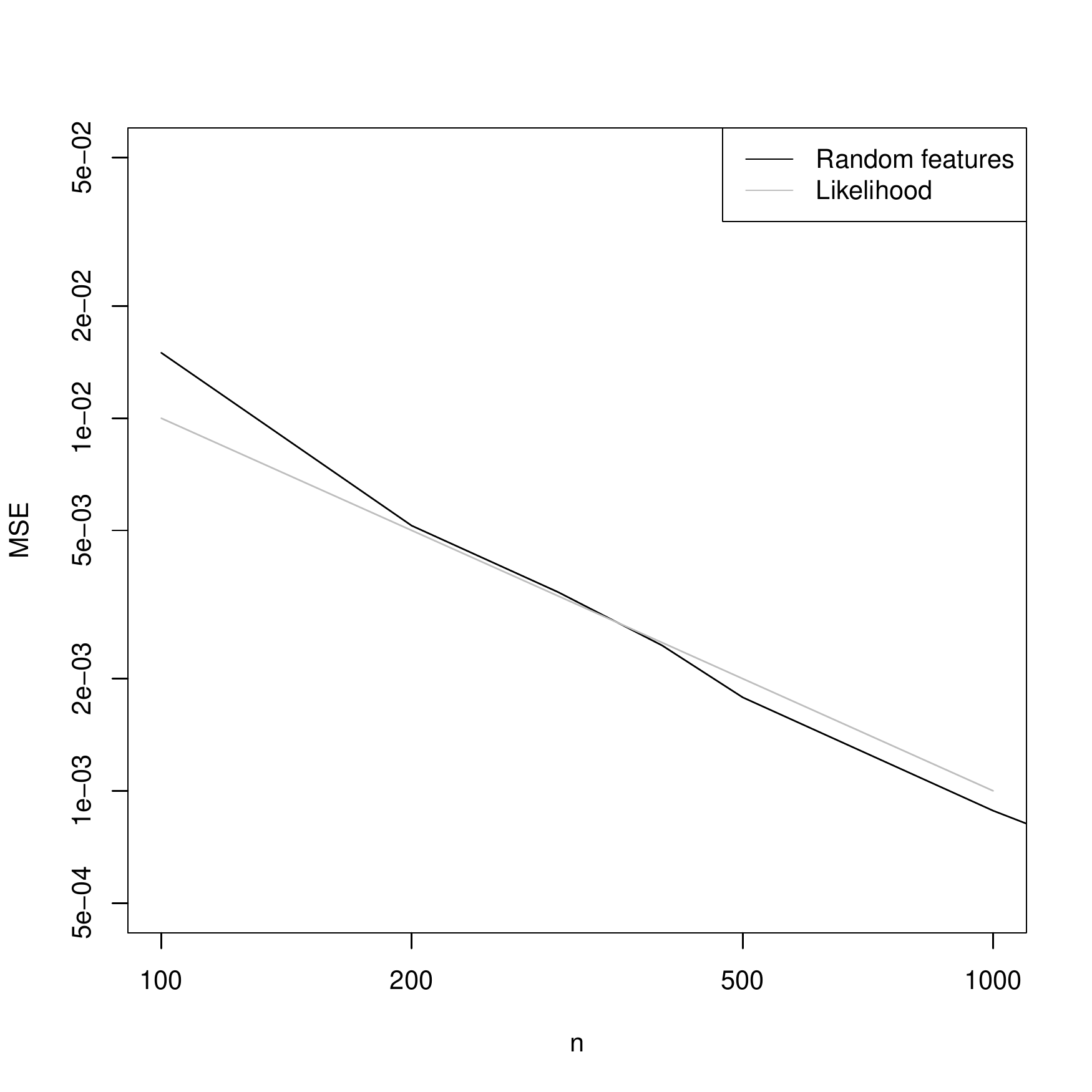} 

\end{knitrout}

\label{fig:mse-of-3-features-for-Gaussian-location}
\caption{MSE of estimating the location parameter of a univariate Gaussian (i.e., $\mu$ in
  $N(\mu, 1)$ family) with 3 random Fourier features and 10 simulations for
  each parameter value considered during the optimization.
  The true $\mu$ is fixed
  at 0 as the sample size is varied (with 100 replicates at each
  sample size.  The grey
  line shows the theoretical MSE of the maximum likelihood estimate, which here is just   $\sigma^2/n = 1/n$.}
\end{figure}

\clearpage

\subsubsection{$t$-Distribution Location Family}

The Gaussian distribution is of course the simplest possible test-case for
estimation.  A natural next step is to consider another location family.
Specifically, I consider the family where $X=\mu+T$, where $T$ is a standard
$t$-distributed random variable with 5 degrees of freedom.  This is
heavy tailed (fifth and higher moments are ill-defined), but still a
one-parameter family where the distribution changes smoothly with the
parameter.

\begin{figure}
\begin{knitrout}\small
\definecolor{shadecolor}{rgb}{1, 1, 1}\color{fgcolor}
\includegraphics[width=\maxwidth]{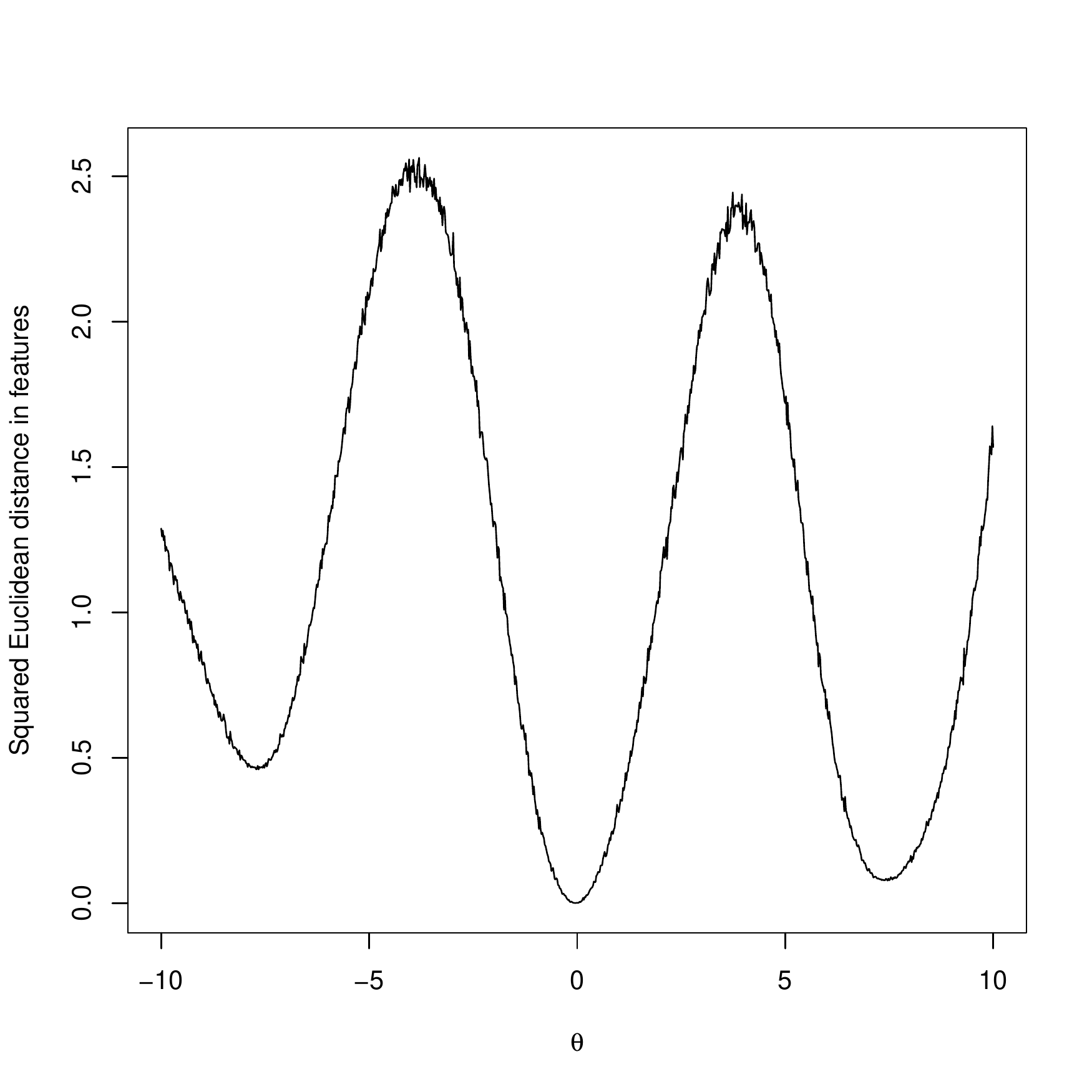} 

\end{knitrout}
\caption{Numerical evaluation of the distance between one sample of a centered $t$
distribution ($n=300$, $df=5$), and $t$ distributions with the same number of degrees of freedom centered at $\theta$.  The distance was evaluated using $10$ random draws per value of $\theta$, and using the {\em same} three random Fourier features used for the Gaussian location examples earlier.}
\label{fig:distance.t.dist}
\end{figure}

\begin{figure}

\begin{knitrout}\small
\definecolor{shadecolor}{rgb}{1, 1, 1}\color{fgcolor}
\includegraphics[width=\maxwidth]{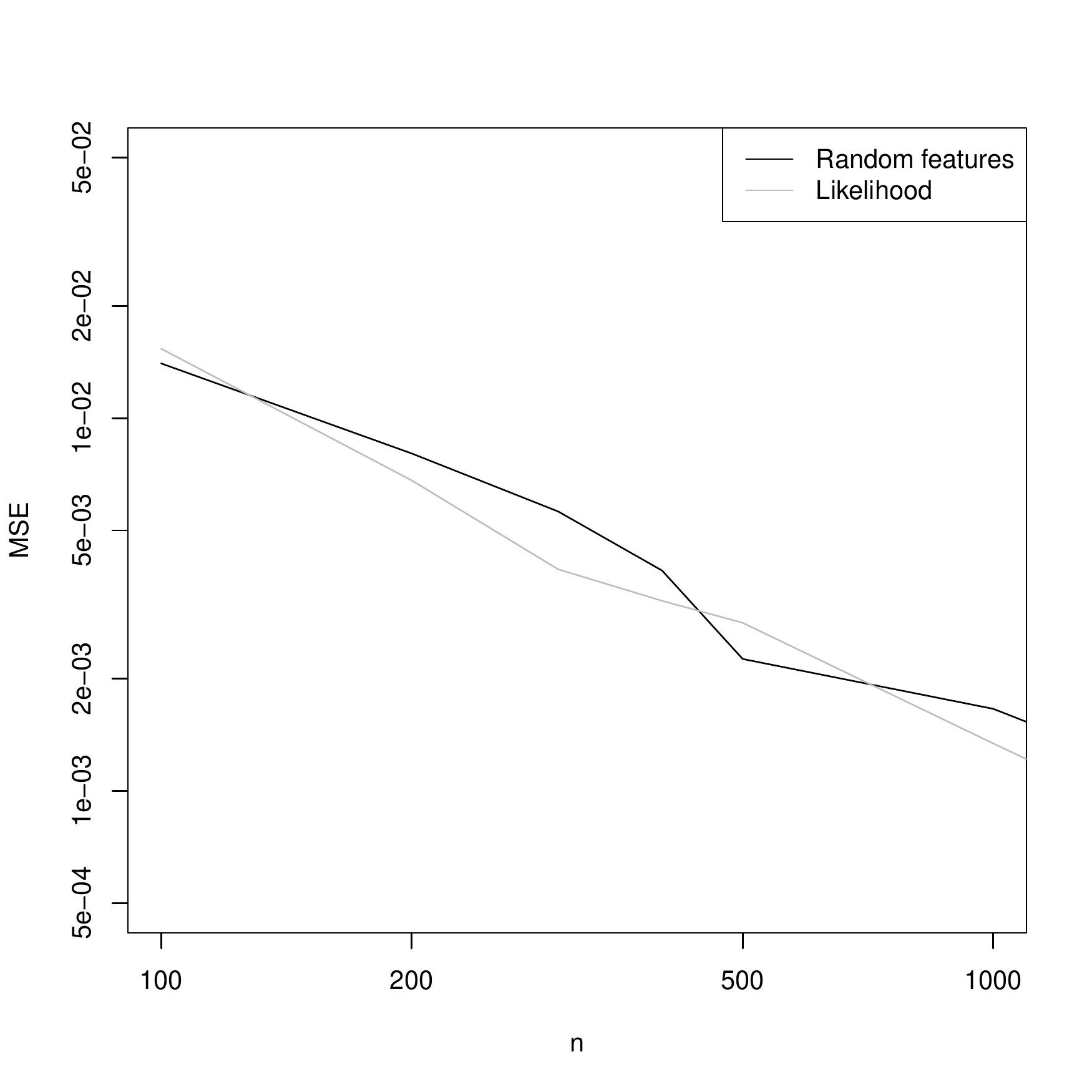} 

\end{knitrout}
\caption{MSEs for estimating the location parameter of a $t$-distribution with 5 degrees of freedom, using distance-minimization on the {\em same} 3 random Fourier features as used for the Gaussian, and for the MLE (evaluated numerically,
  using the \texttt{fitdistr} function from the \texttt{MASS} library \citep{Venables-Ripley}). Results are averages over 100 trials, with 10 simulations per considered parameter value.}
\label{fig:t.dist.location.RFM.est}
\end{figure}

Using the {\em same} three random Fourier features as in the Gaussian example
still manifestly gives an embedding (Figure
\ref{fig:distance.t.dist}). Minimizing distance in those features gives
estimates which are not much worse than the MLE\footnote{In this case the MLE,
  too, is found by numerical optimization, as implemented in the
  \texttt{fitdistr} function in the \texttt{MASS} package
  \citep{Venables-Ripley}.} (Figure \ref{fig:t.dist.location.RFM.est}).

\clearpage

\clearpage

\clearpage

\subsubsection{Estimating a Dynamical Systems from Its Invariant Distribution}

We don't actually need the machinery of simulation-based inference to estimate
the location parameters of bell curves, even heavy-tailed ones.  I thus turn to
a much more challenging example, namely the {\bf logistic map}, a deterministic dynamical system defined by
\begin{equation}
 S_{t+1} = 4r S_{t} (1-S_{t})
\end{equation}
where both the state variable $S_t \in [0,1]$, and the dynamical parameter $r$
are $\in [0,1]$.  The observable, above $X_n$, will be the whole sequence or
trajectory $(S_1, S_2, \ldots S_n)$, so $\mathcal{X}_n = [0,1]^n$.

\begin{figure}
\begin{knitrout}\small
\definecolor{shadecolor}{rgb}{1, 1, 1}\color{fgcolor}
\includegraphics[width=\maxwidth]{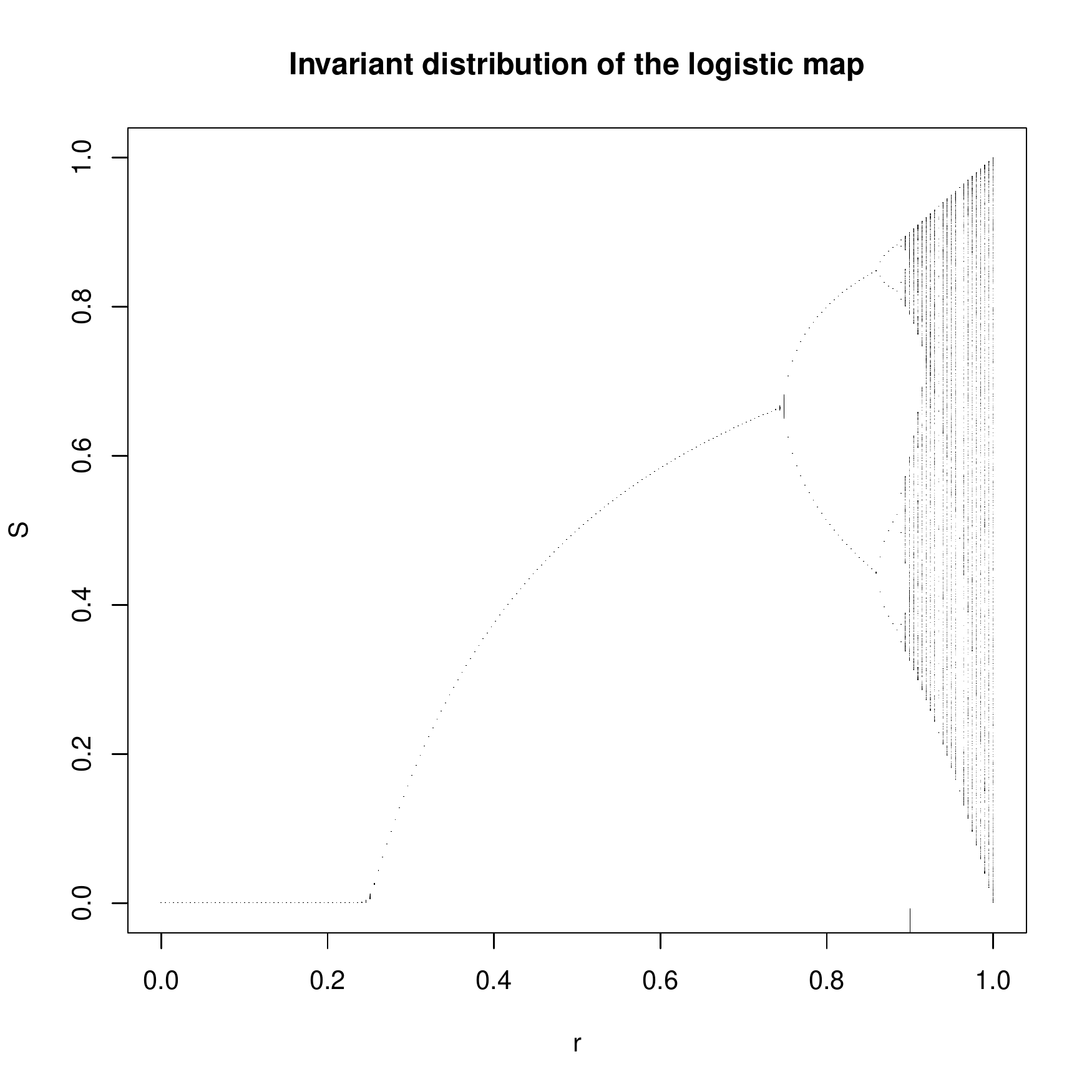} 

\end{knitrout}
\caption{A (partial) display of the invariant distribution of the logistic map as a function of $r$.  (It's only partial because the distribution within the vertical bands is non-uniform.)  The tick on the horizontal axis at $r=0.9$ indicates the true value of the parameter used in the estimation experiments that follow, though additional simulations (not shown) show very similar behavior for other values of $r$.}
\label{fig:logisticmap.invariant.dist}
\end{figure}

For small values of $r$, the sole fixed point is $S=0$ and every trajectory
approaches it, so the only invariant distribution is the point mass at 0.
Otherwise, every $r$ has its own invariant distribution over $[0,1]$ (Figure \ref{fig:logisticmap.invariant.dist}).  If $S_1$
is drawn from this invariant distribution, then the trajectory $S_1, S_2,
\ldots$ is the realization of a stationary, and indeed ergodic, Markov process,
and every $r$ has a unique natural ergodic distribution over trajectories.  (Even if $S \sim \mathrm{Unif}(0,1)$, the trajectory very quickly approaches stationarity.)  For
sufficiently large values of $r$, the system may be {\bf chaotic}, meaning that
it is both ergodic {\em and} shows sensitive dependence on initial conditions.
One such parameter value is $r=0.9$, which is what will be used in
the experiments below, though the behavior of the estimation method at other
parameter values is quite similar, whether or not the dynamics are
chaotic\footnote{This terse summary does no justice to the quite intricate
  mathematical theory which has been built up around the logistic map.  For an
  introduction, \citet{Devaney-first-course} is still extremely valuable.}.

Note that in this case, it's hard to see how we could use the method of maximum
likelihood as a baseline.  Since $S_1$ is observed, for a particular $r$, the
trajectory $(S_1, S_2, \ldots S_n)$ is either impossible (likelihood 0) or
mandatory (likelihood 1).  While it's true that the correct $r$ will uniquely
maximize this likelihood, actually doing the optimization would be challenging
to say the least.  (Alternatively, of course, the ratio
$\frac{S_2}{S_1(1-S_1)}$ is enough to determine $r$ exactly.)  Various {\em ad
  hoc} modifications are of course possible (e.g., saying we only observe
whether or not each $S_t$ is in a bin of some narrow width), but it seems
better to just admit that this is a domain where the method of maximum
likelihood fails us.

\begin{figure}
\begin{knitrout}\small
\definecolor{shadecolor}{rgb}{1, 1, 1}\color{fgcolor}
\includegraphics[width=\maxwidth]{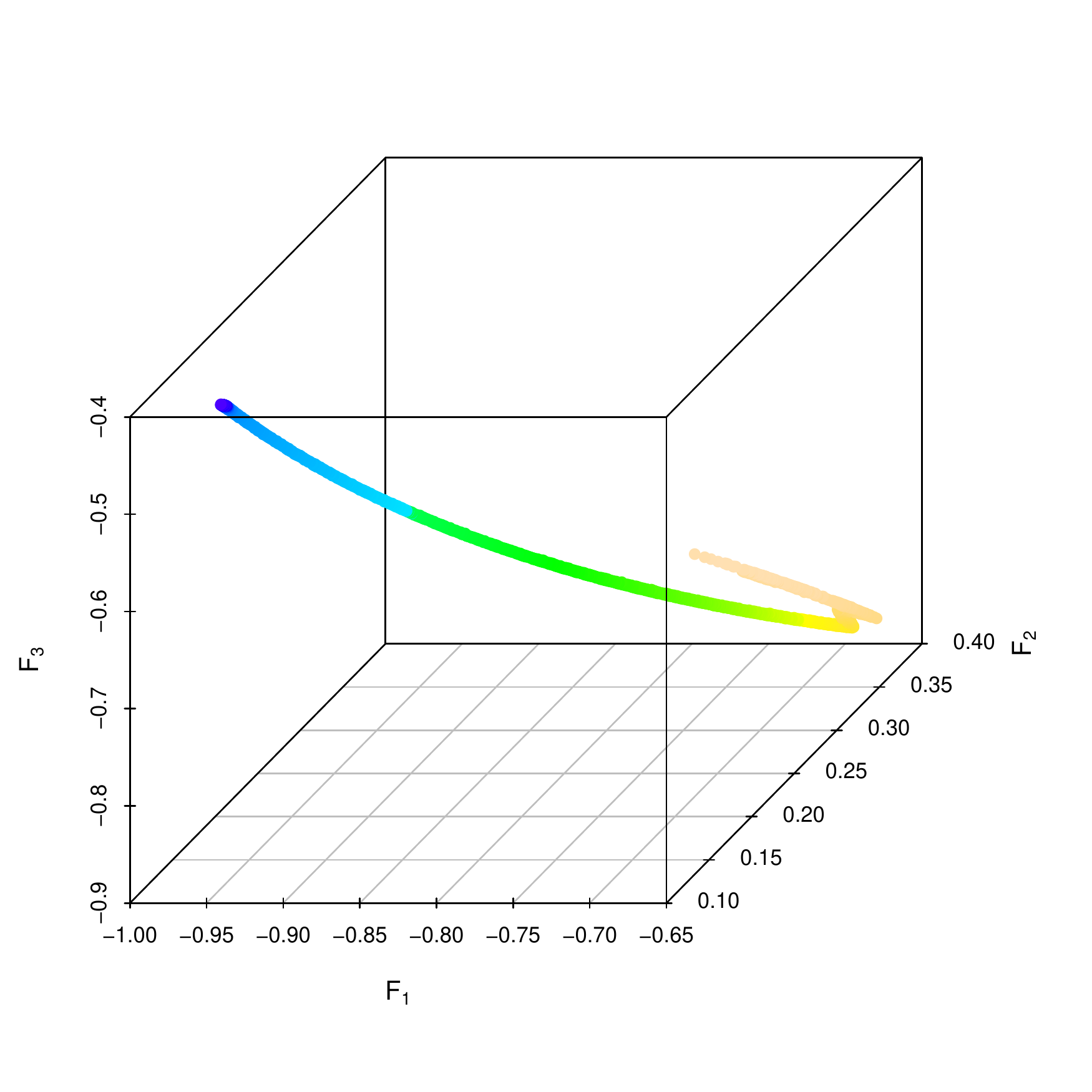} 

\end{knitrout}
\caption{Values of the {\em same} univariate Fourier features used for the Gaussian example, evaluated along random trajectories of the logistic map, as a function of the logistic map parameter $r$.  Features are evaluated by averaging 10 trajectories of length 100 each, with initial conditions chosen randomly in $[0,1]$.  (Strictly speaking, then, these are not stationary time series.)}
\label{fig:logistic.univariate.feature.mapping}
\end{figure}

As before, I will begin by showing that the values of 3 ($=2\ModelDim+1$)
random Fourier features will uniquely and smoothly identify the parameter $r$.
Figure \ref{fig:logistic.univariate.feature.mapping} does this using the {\em
  same} three features previously used in the IID location examples, to
emphasize that we need to understand very, very little about the underlying
model's behavior in order to pick adequate features.

\clearpage

\clearpage

\clearpage

\begin{figure}
\begin{knitrout}\small
\definecolor{shadecolor}{rgb}{1, 1, 1}\color{fgcolor}
\includegraphics[width=\maxwidth]{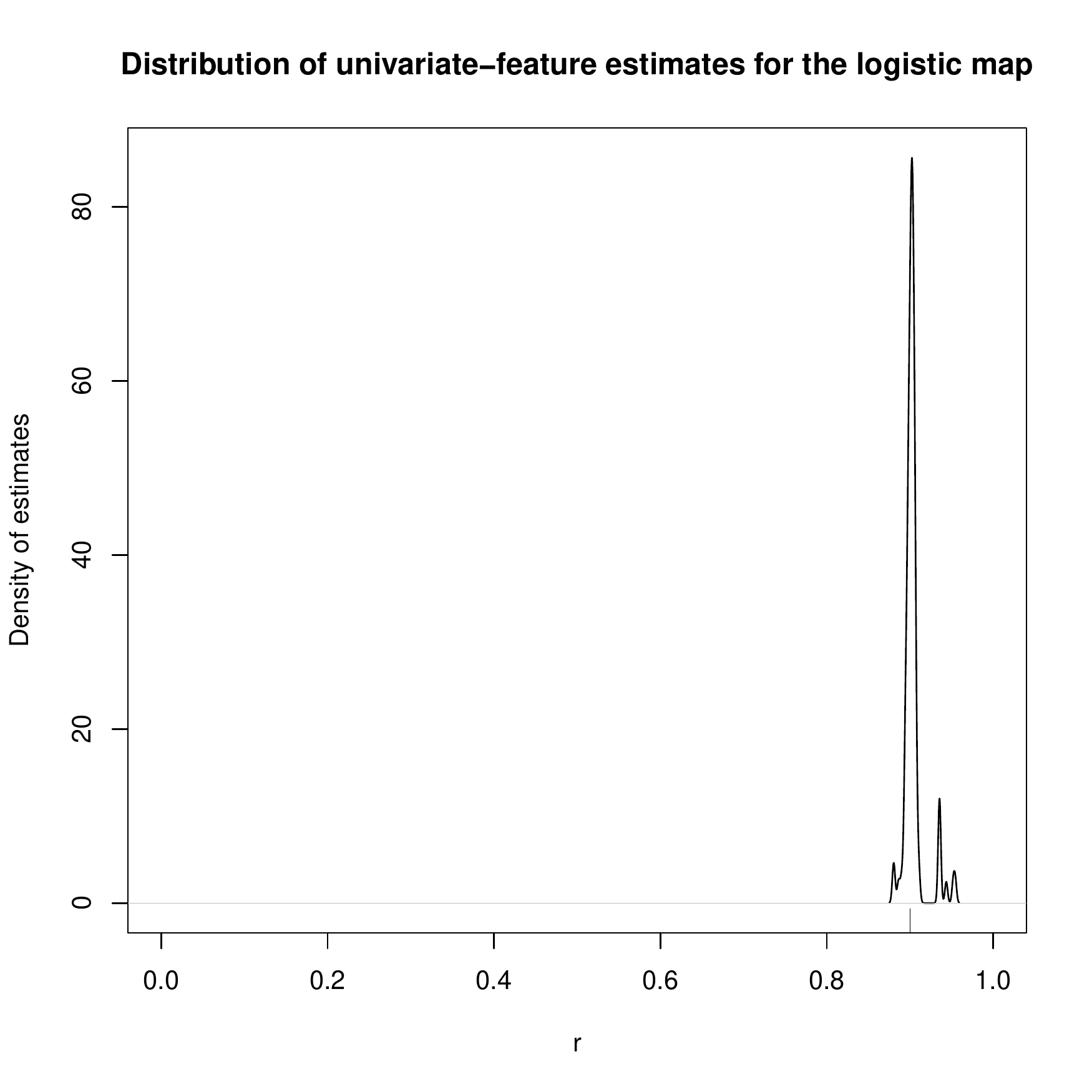} 

\end{knitrout}
\caption{Density of random-feature matching estimates (kernel-smoothed) for
         the parameter $r$ of the logistic map from time series of length 100 and 10 simulations
       per parameter value, minimizing distance with respect to the same three univariate Fourier used for the location-family examples above.  The true parameter was $r=0.9$ for all trials.  The ``data'' trajectory was randomly regenerated for each of the 100 simulation trials, but the random features were kept constant.  (Additional experiments, not included here, show that randomly regenerating the features for each trial broadens the distribution only slightly.)  Note that subsequent figures will zoom in on the horizontal axis.}
\end{figure}



\clearpage

\subsubsection{Estimating a Dynamical Systems from Bivariate Distributions}

While, as mentioned, all large-enough values of $r$ lead to distinct invariant
distributions over $[0,1]$, there is clearly some loss of information in going
from whole trajectories to univariate random features.  Using bivariate random
Fourier features (as defined above) seems like a natural compromise,
particularly if the models we're entertaining are deterministic
dynamical systems.

\begin{figure}
\begin{knitrout}\small
\definecolor{shadecolor}{rgb}{1, 1, 1}\color{fgcolor}
\includegraphics[width=\maxwidth]{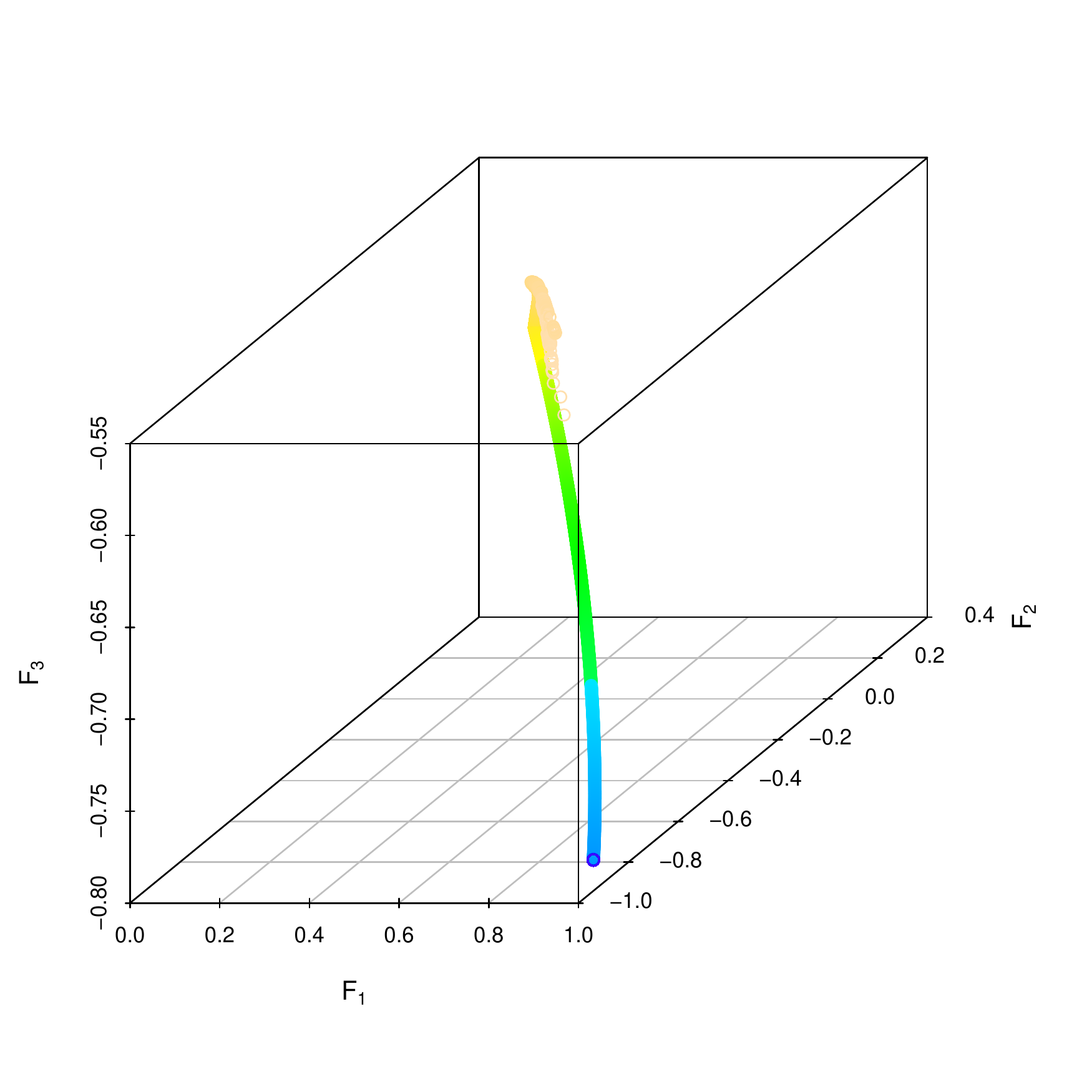} 

\end{knitrout}
\caption{Identifiability of the logistic map parameter from bivariate random
  Fourier features, but otherwise as in Figure
  \ref{fig:logistic.univariate.feature.mapping}.}
\label{fig:ident.from.two.point.fourier.features}
\end{figure}

Again, numerically there is indeed an embedding using three random
bivariate Fourier features (Figure
\ref{fig:ident.from.two.point.fourier.features}).  That being the case,
I turn to estimation.

\clearpage

\begin{figure}
\begin{knitrout}\small
\definecolor{shadecolor}{rgb}{1, 1, 1}\color{fgcolor}
\includegraphics[width=\maxwidth]{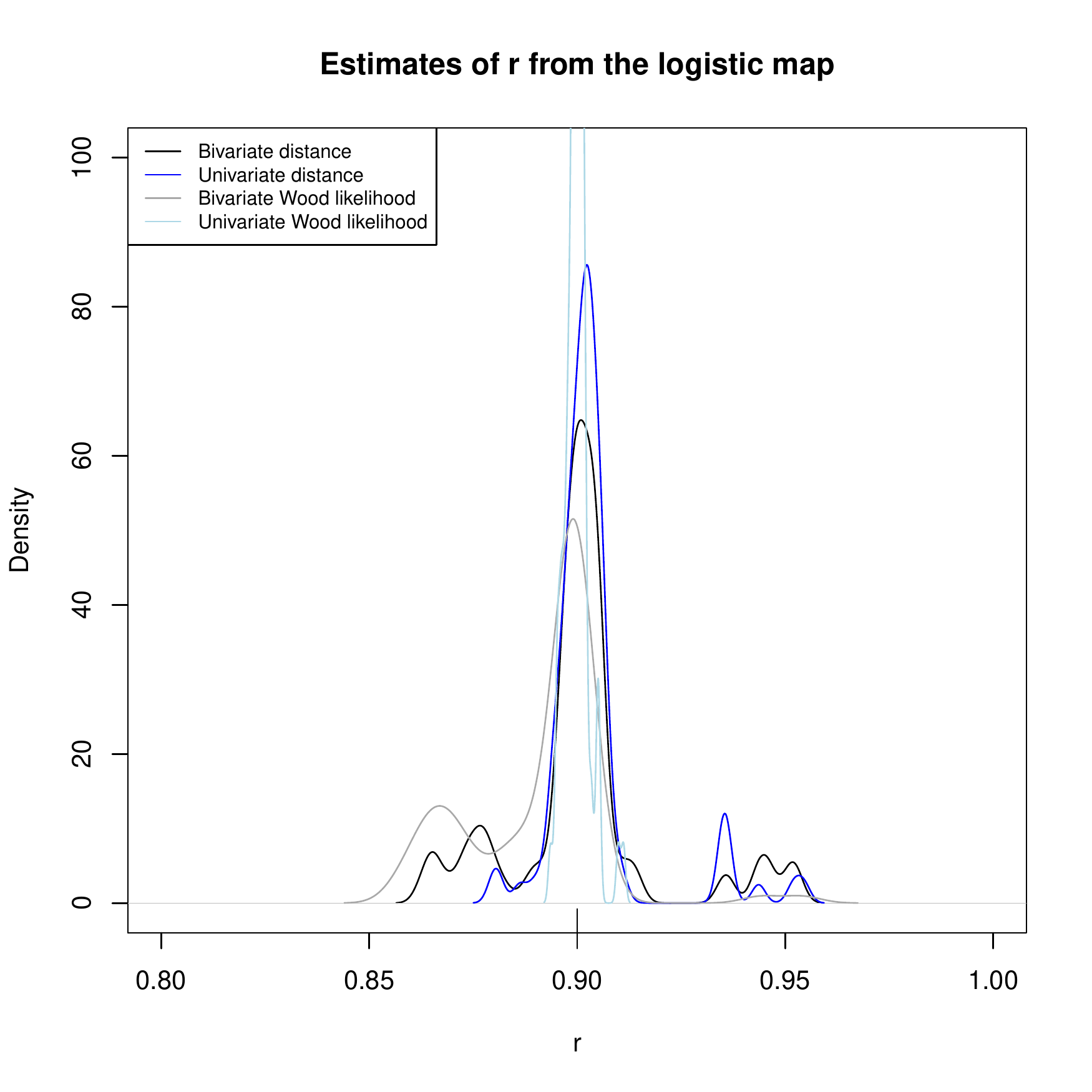} 

\end{knitrout}
\caption{Distribution of estimates for the logistic map parameter value using 3
  bivariate (black) and univariate (grey) random Fourier features, and $s=10$
  simulations per parameter value in optimization.  All time series are of
  length $100$, and the true parameter value,
  $0.9$, is marked by a tick on the axis.  (The Wood likelihood
  estimates using univariate features are very concentrated around
  $r=0.9$, and so the vertical scale of this plot is truncated to
  show the shape of the other distributions.)}
\label{fig:eval-two-point-basic-estimates}
\end{figure}

\clearpage

\begin{figure}
\begin{knitrout}\small
\definecolor{shadecolor}{rgb}{1, 1, 1}\color{fgcolor}
\includegraphics[width=\maxwidth]{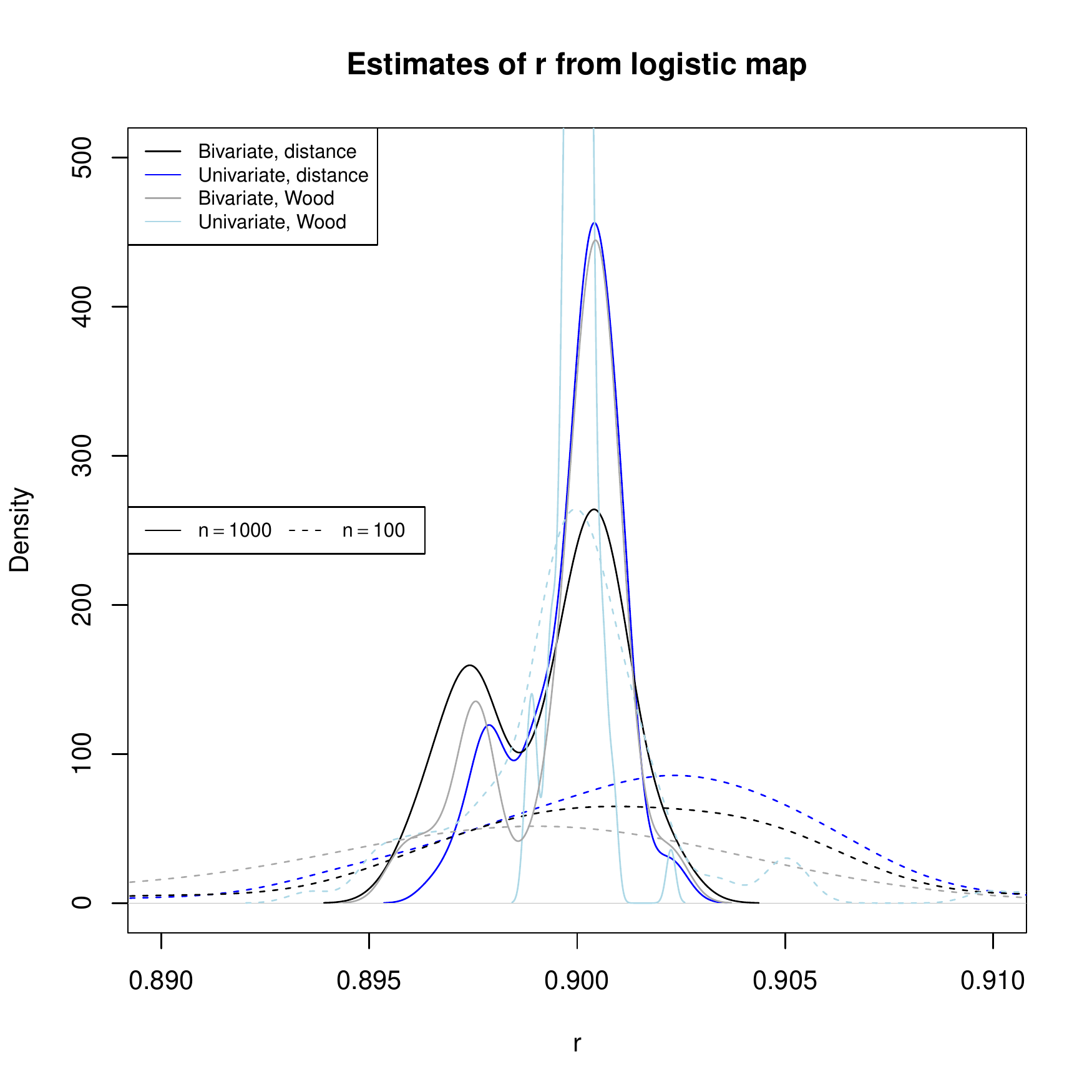} 

\end{knitrout}
\caption{Distribution of estimates of $r$ from the logistic map
  (kernel-smoothed), comparing univariate vs.\ bivariate features, simple
  distance minimization vs.\ maximizing the Wood likelihood, and two sample
  sizes, all as in Figure \ref{fig:eval-two-point-basic-estimates}.  Notice
  that, compared to that figure, the horizontal axis has been zoomed in around
  the true parameter value $r=0.9$, and, again, the vertical axis
  cuts off the density of the most-concentrated estimator.}
\end{figure}

\clearpage

\subsection{Logistic Map Observed Through Noise}

So far, I have only dealt with one-parameter families, though the logistic
map is a rather tricky one.  To show that
random-feature-matching doesn't just work because of some quirk of
one-dimensional families, I will now consider a {\em two} parameter family,
which is actually a hidden Markov model, namely the logistic map observed
through Gaussian noise.  Specifically, the model specification is
\begin{eqnarray}
  S_{t+1} | S_t & = & 4 r S_t (1-S_t)\\
  Y_t | S_t & \sim & \mathcal{N}(S_t, \sigma^2)
\end{eqnarray}
with only $Y_t$ being observable, so $X_n = (Y_1, Y_2, \ldots Y_n)$.  That is,
in the hidden layer, $S_t$ follows the deterministic logistic map, but we do
not get to see it face to face, but only through the distortion of additive
Gaussian noise of variance $\sigma^2$.  There are thus two parameters to
estimate, $r$ and $\sigma^2$, so we require 5 random features.

\begin{figure}
\begin{knitrout}\small
\definecolor{shadecolor}{rgb}{1, 1, 1}\color{fgcolor}
\includegraphics[width=\maxwidth]{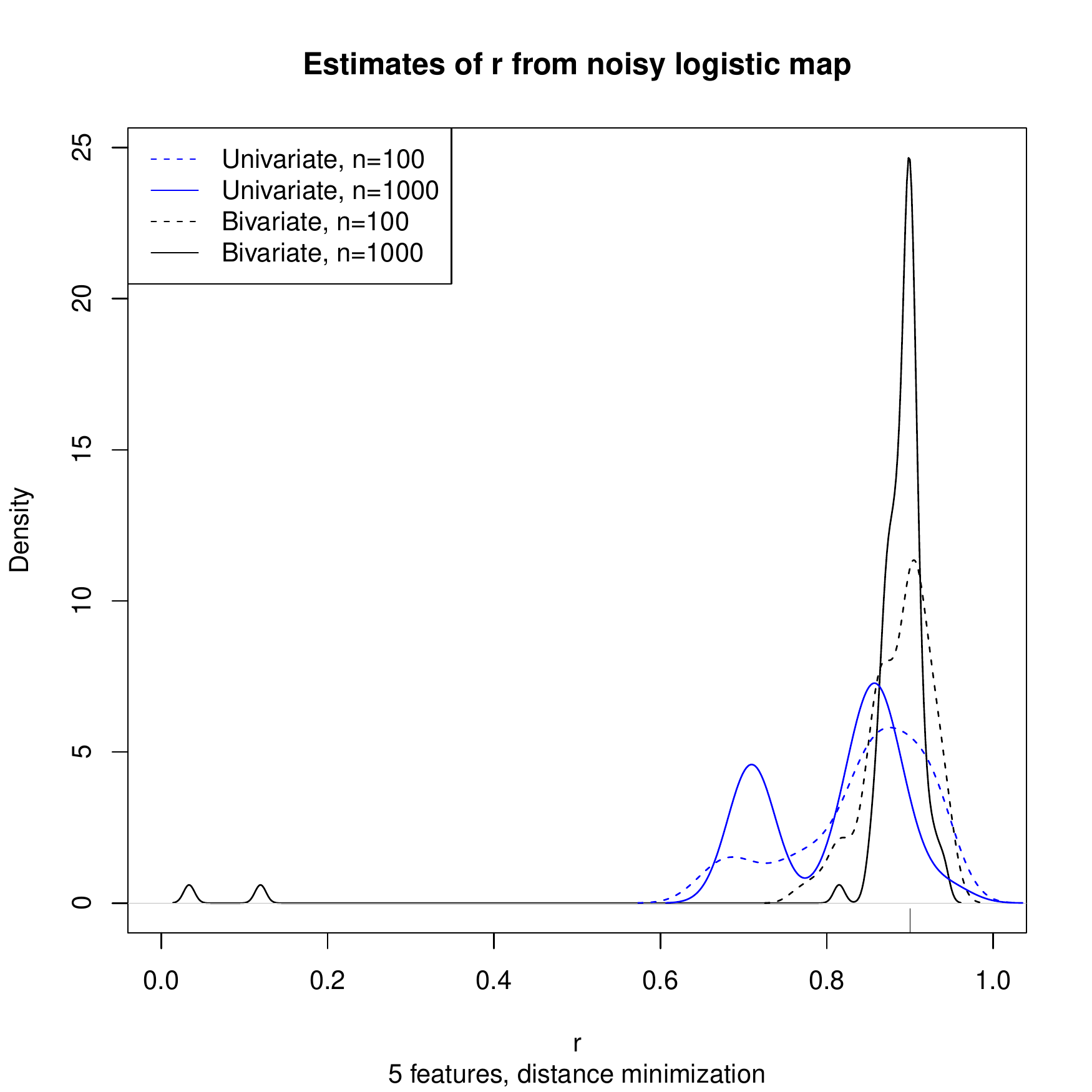} 

\end{knitrout}
\caption{Density of estimates of $r$ from the noisy logistic map (kernel
  smoothed), showing univariate vs.\ bivariate features (blue vs. black) and
  $n=100$ vs. $n=1000$ (solid
  vs. dashed lines).  10 simulations were used per parameter value when
  optimizing.  The true parameter value was held fixed at $r=0.9$.}
\label{fig:display.noisy.logistic.estimates.7.r}
\end{figure}

\begin{figure}
\begin{knitrout}\small
\definecolor{shadecolor}{rgb}{1, 1, 1}\color{fgcolor}
\includegraphics[width=\maxwidth]{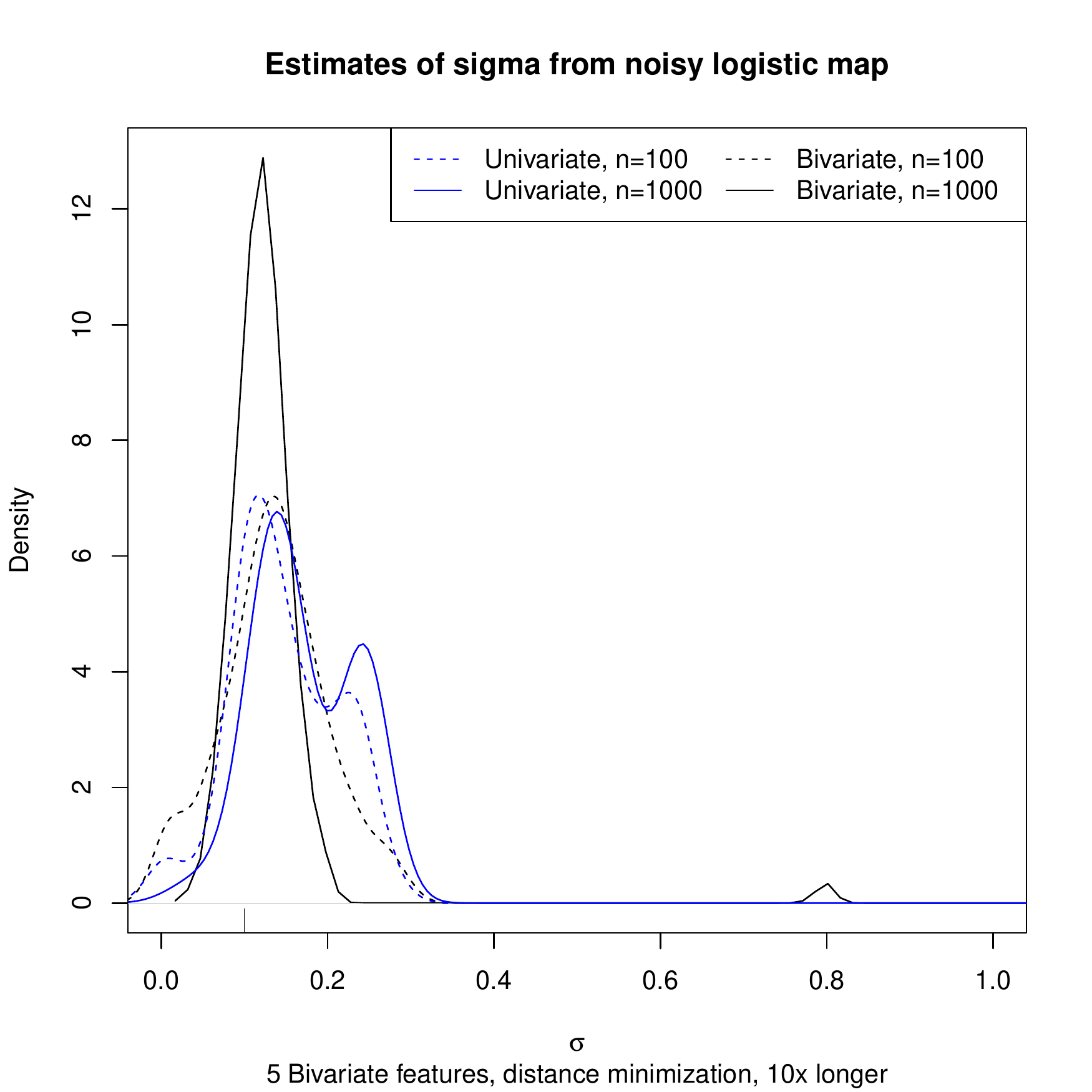} 

\end{knitrout}
\caption{As in Figure \ref{fig:display.noisy.logistic.estimates.7.r}, but for estimates of $\sigma$ (not $\sigma^2)$; the true parameter value was held fixed at $\sigma=0.1$.}
\label{fig:display.noisy.logistic.estimates.7.sigma}
\end{figure}

Because I despair of making a five-dimensional plot, I will skip the visual
display of embedding, and just go straight to reporting distributions of
estimates (Figures \ref{fig:display.noisy.logistic.estimates.7.r} and
\ref{fig:display.noisy.logistic.estimates.7.sigma}).  Unsurprisingly, estimates
of $r$ are less precise, at equal sample sizes, than in the noise-free case,
but the estimates are also, visibly, converging on the correct parameter
values.








\clearpage

\subsection{Notes on These Experiments}

The last sub-section, on the logistic map seen through noise, shows that we can
get quite good estimates using univariate and bivariate random Fourier features
on a hidden Markov model.  (This indicates that random feature matching doesn't
just work on Markov processes.)  It is natural to wonder whether the choice of
univariate, bivariate, or higher-order features could be automated.
Information-theoretic \citep{Marton-Shields,
  Steif-joint-distributions-of-random-fields} and learning-theoretic
\citep{McDonald-density-estimation-for-growing-dimension} results on
estimating joint distributions both suggest that with $n$ observations, it is
feasible to non-parametrically estimate the distribution of blocks of whose
length scales like $\log{n}$, but no faster.  Of course we are dealing with
parametric families of distributions, so a faster rate of block-growth with $n$
may be feasible, but $\log{n}$ should be safe.

These experiments have also been limited to stationary models\footnote{Strictly
  speaking, because the logistic map examples are not started from the
  invariant distribution, they are not stationary, but they are asymptotically
  mean-stationary \citep{Gray-ergodic-properties}, and they rapidly approach
  the stationary limit.}  Suitable choices of random features for
non-stationary processes, which do not prejudge the form of the
non-stationarity, will be reported elsewhere.

\clearpage

\section{Summary}

The argument of this manuscript is that when we want to do simulation-based
inference on a model with $\ModelDim$ parameters, we will generally be able to
do so using $2\ModelDim+1$ random nonlinear features, i.e., functions of the
data, chosen independently of the data, and indeed of the model.  The theory
sketched above suggests that these features should come from a class of
functions whose linear combinations are dense in the space of bounded,
continuous test functions on the sample space.  They should also be functions
whose sample- or time- averages converge (rapidly) on expectation values.
Random Fourier features therefore suggest themselves, but are by no means
required.  The numerical experiments reported above show that matching random
features can be competitive, statistically, with maximum likelihood, and
delivers good results in some situations where likelihood-based inference is
scarcely feasible.

A more detailed treatment of the many theoretical and implementation questions
raised by these preliminary results is in preparation.  In the meanwhile, these
results suggest that simulation-based inference can be made (much more nearly)
automatic through matching random features, at little statistical cost.

\paragraph*{Acknowledgments} A. E. Owen offered invaluable moral and
intellectual support, and soundly vetoed multiple bad names for the technique.
I wish to acknowledge the grant review panel which rejected a proposal based on
this idea, thereby providing crucial motivation.

\bibliography{locusts}
\bibliographystyle{crs}

\end{document}